\documentclass[useAMS,usenatbib,usegraphicx,11pt]{article}
\pdfoutput=1 
\usepackage{jcappub}
\usepackage[english]{babel}
\usepackage{xcolor}
\usepackage{soul}
\usepackage{multirow}
\usepackage{colortbl}
\usepackage{units}
\usepackage{mathrsfs}
\usepackage[super]{nth}
\usepackage{bm}
\usepackage[nameinlink,noabbrev]{cleveref}
\usepackage{subfigure}
\usepackage[hang, small, bf]{caption}
\usepackage{sidecap}
\usepackage{afterpage}

\newcommand{\gr}{$\gamma$-ray}
\newcommand{\grs}{$\gamma$-rays}

\newcommand{\beq}{\begin{equation}}
\newcommand{\eeq}{\end{equation}}
\newcommand{\balign}{\begin{align}}
\newcommand{\ealign}{\end{align}}

\newcommand{\lcdm}{{\ifmmode \Lambda{\rm CDM} \else $\Lambda{\rm CDM}$\fi}}
\newcommand{\mchi}{\ensuremath{m_{\chi}}}

\newcommand{\healpix}{{\tt HEALPix}}

\newcommand{\degs}{{^{\circ}}}
\newcommand{\nside}{{N_{\rm side}}}

\newcommand{\sigmav}{\ensuremath{\langle \sigma v\rangle}}

\newcommand{\dd}{\ensuremath{\mathrm{d}}}
\newcommand{\fermi}{{\em Fermi}-LAT}

\newcommand{\npix}{\ensuremath{N_{\rm pix}}}
\newcommand{\nevents}{\ensuremath{N_{\rm ev}}}
\newcommand{\nsig}{\ensuremath{N_{\rm DGRB}}}
\newcommand{\nbck}{\ensuremath{N_{\rm bck}}}
\newcommand{\fsky}{\ensuremath{f_{\rm sky}}}

\graphicspath{{./}}

\crefname{subsection}{\S}{\S}
\crefname{equation}{Eq.}{Eq.}

\title{\boldmath Observing small-scale \gr{} anisotropies with the Cherenkov Telescope Array}

\author[a,b,c]{M. H\"utten}
\author[c]{and G. Maier}
\affiliation[a]{Max-Planck-Institut f\"ur Physik, F\"ohringer Ring 6,  D-80805 M\"unchen, Germany}
\affiliation[b]{Humboldt-Universit\"{a}t zu Berlin, Newtonstra{\ss}e 15, D-12489 Berlin, Germany}
\affiliation[c]{DESY, Platanenallee 6, D-15738 Zeuthen,  Germany}
\emailAdd{mhuetten@mpp.mpg.de}
\emailAdd{gernot.maier@desy.de}

\abstract{
Disentangling the composition of the diffuse \gr{} background (DGRB) is a major challenge in \gr{} astronomy. It is presumed that at the highest energies, the DGRB is dominated by relatively few, still unresolved point sources. This conjecture has recently been  supported by the measurement of small-scale anisotropies in the DGRB by the \textit{Fermi} Large Area Telescope (LAT) up to energies of $\unit[500]{GeV}$. We show how such anisotropies can be searched for with the forthcoming Earth-bound Cherenkov Telescope Array (CTA) up to the TeV range. We investigate different observation modes to analyse CTA data for small-scale anisotropies and propose the projected  extragalactic large-area sky survey as the most promising data set. Relying on an up-to-date model of the performance of the southern CTA, we find that CTA will be able to probe anisotropies in the DGRB from unresolved point sources at a relative amplitude of $C_{\rm P}^I/I^2_{\rm DGRB}\gtrsim 4\times \unit[10^{-3}]{sr}$ at energies above $\unit[30]{GeV}$ and angular scales $\lesssim 1.5\degs$. 
Such DGRB anisotropies have not yet been ruled out by the \fermi{}. The proposed analysis would primarily clarify the contribution from blazars and misaligned active galactic nuclei to the very-high-energy regime of the DGRB, as well as provide insight into dark matter annihilation in Galactic and extragalactic density structures. Finally, it constitutes a measurement with complementary systematic uncertainties compared to the \fermi{}.
}


\begin{document}
\maketitle
\flushbottom


\section{Introduction}
\label{sec:intro}

The last decade has marked a major breakthrough in \gr{} astronomy. The \textit{Fermi} Large Area Telescope (LAT, \cite{2009ApJ...697.1071A}), operating  since 2008, has resolved thousands of astrophysical \gr{} sources above energies of \unit[100]{MeV} \cite{2015ApJS..218...23A}. Ground-based instruments like H.E.S.S., MAGIC, and VERITAS revealed over a hundred sources emitting \grs{} in the very-high-energy (VHE) regime beyond \unit[100]{GeV}.\footnote{\url{http://tevcat2.uchicago.edu/} \cite{2008ICRC....3.1341W}} The presence of diffuse Galactic emission along the Milky-Way plane and its origin in cosmic-ray interaction with the interstellar gas and radiation fields in the Galaxy are well established \cite{1972ApJ...177..341K,1975ApJ...198..163F,2012ApJ...750....3A}. The radiation processes giving rise to the diffuse \textit{Fermi}-bubbles \cite{2010ApJ...724.1044S} and Loop-I structure \cite{2009arXiv0912.3478C}, extending far above the Galactic disk, are still under debate. Nonetheless, the \fermi{} has  detected millions of GeV photons, which cannot be attributed neither to resolved localised sources or to known diffuse structures \cite{2010PhRvL.104j1101A,2015ApJ...799...86A}: These \grs{} compose the so-called diffuse \gr{} background (DGRB)\footnote{The DGRB is synonymously denoted as \textit{isotropic \gr{} background} (IGRB), \textit{extragalactic diffuse \gr{} background} (EDGB), or \textit{unresolved \gr{} background} (UGRB) throughout the literature.} and, by definition, comprise all \grs{} outside the Galactic plane with almost isotropic distribution and not assigned to known \gr{} sources on the sky \citep{2007AIPC..921..122D,2015PhR...598....1F}. Unresolved, far-away extragalactic \gr{} sources, distributed isotropically on the sky, definitely constitute a major fraction of the DGRB \cite{2014JCAP...09..043T,2014ApJ...780..161D,2015PhRvD..91l3001D,2015ApJ...800L..27A,2016PhRvL.116o1105A,2018ApJ...856..106D}. However, also unresolved sources of Galactic origin at high latitudes \cite{2011MNRAS.415.1074S,2014ApJ...796...14C} and unaccounted large-scale Galactic diffuse emission \cite{2015PhR...598....1F} might contribute to the DGRB. Annihilation of dark matter (DM) particles in the Milky Way halo and throughout the cosmic web might provide an additional, still unidentified \gr{} source \cite{1978ApJ...223.1015G,1978ApJ...223.1032S,2018JCAP...02..005H}. The relative contributions of numerous Galactic and extragalactic source classes to the DGRB are currently under debate \cite{2015PhR...598....1F}. 

The prevalence of extragalactic sources in the DGRB is further supported by the recent observation of its spectral softening below $\unit[1]{TeV}$ \cite{2015ApJ...799...86A}, in agreement with an  attenuation due to pair-production losses from interaction with the extragalactic background light (EBL, \cite{2015ApJ...805...33K}). If the DGRB is of mostly extragalactic origin, it must necessarily become more anisotropic at higher energies. Any known physical process generating \grs{} can only occur in the surroundings of galaxies or galaxy clusters, which  appear virtually point-like on the sky for current \gr{} telescopes. The \grs{} arriving from these far-away sources are redshifted by the cosmic expansion and attenuated by the EBL, with the latter effect being stronger the higher the \gr{} energy \cite{2008A&A...487..837F}. For these reasons, only relatively close or spectrally hard objects are expected to emit \grs{} that are observable at VHE energies, independent of the specific production mechanism at source. This conjecture is in agreement with the identified extragalactic sources in the third \fermi{} full-sky catalogue of hard-spectra \gr{} sources (3FHL) \cite{2016ApJS..222....5A} and the angular directions and redshifts of known VHE \grs{} emitters.

Focusing on such anisotropic signatures of the DGRB to decipher its origin, the  approach of an angular power spectrum (APS) analysis was first applied to data from the \fermi{} several years ago. An anisotropic component to the DGRB on angular scales $\lesssim 2^{\circ}$ (``small-scale anisotropies'') was in fact observed  \cite{2012PhRvD..85h3007A}. The amplitude of these anisotropies was found compatible with being constant on all investigated  angular scales, suggesting that the APS signal is caused by unclustered point sources. 

A recent analysis up to $\unit[500]{GeV}$ based on more than six years of \fermi{} data   confirmed this finding and indicates a spectral hardening of the underlying sources with increasing energy \cite{2016PhRvD..94l3005F}. The detected APS is in agreement with the prediction for unresolved blazars \cite{2007PhRvD..75f3519A,2012PhRvD..86f3004C} and misaligned active galactic nuclei (AGN) \cite{2014JCAP...11..021D}. However, it has been suggested that also Galactic sources, in particular high-latitude millisecond pulsars  \cite{2011MNRAS.415.1074S}, or relic annihilation of Galactic and extragalactic DM \cite{2006PhRvD..73b3521A,2008JCAP...10..040S,2009PhRvD..80b3518F,2011MNRAS.414.2040C,2013MNRAS.429.1529F,2013PhRvD..87l3539A,2014NIMPA.742..149G,2015MNRAS.447..939L} might significantly contribute to small-scale DGRB  anisotropies.

Further knowledge about the contributors to the DGRB can be obtained by probing its APS in the VHE regime. Due to the large cosmic-ray background, it is highly challenging to measure the absolute intensity of diffuse \grs{} with Earth-bound \gr{} detectors. However, the measurement of \gr{} \textit{anisotropies} from ground may be much less constrained by the background. It was found by  \cite{2014JCAP...01..049R} that current instruments  and their available data sets provide only limited capabilities for probing VHE anisotropies in the DGRB. However, with the future Cherenkov Telescope Array (CTA, \cite{2013APh....43....3A}), a relatively large field of view (FOV) of a ground-based instrument will be combined with the best angular resolution ever achieved in \gr{} astronomy. With CTA, the authors of \cite{2014JCAP...01..049R} have shown that a promising sensitivity to small-scale VHE \gr{} anisotropies can be reached. In the last years, optimization studies for the array layout and detailed studies on the expected performance of CTA have been performed, and  dedicated observing plans for the first decade of operation have been drawn \cite{2017arXiv170907997C}. Therefore, this paper presents a refined assessment of CTA's capability to resolve small-scale anisotropies in the DGRB.

This article is organised as follows: In \cref{sec:aps_analysis}, we introduce the concept of angular power spectra  and describe our likelihood-based analysis method of event data APS. In \cref{sec:cta}, we present the instrumental model for the CTA event sampling (\cref{sec:cta_instrumentcharact}) and outline our setup of a CTA extragalactic survey (\cref{sec:cta_survey}). We investigate the APS  characteristic of the CTA cosmic-ray background and argue to prefer data from a shallow large-area survey for a study of \gr{} anisotropies  (\cref{sec:aps_analysis_background}). \Cref{sec:results} then presents our analysis and  results on the CTA sensitivity to small-scale anisotropies. Our findings are finally discussed in \cref{sec:discussion} with respect to existing data and expected DGRB anisotropies caused by different source classes. We conclude in \cref{sec:conclusions}.


\section{Likelihood-based analysis of event data angular power spectra (APS)}
\label{sec:aps_analysis}

We quantify the anisotropies in the DGRB  by  decomposing  the spatial distribution of \gr{}-like events into its auto-correlation angular power spectrum (APS), $C_{\ell}$. A square-integrable function $I(\vartheta, \varphi) = I(\vec{k})$ on the sphere can be expressed as a linear combination of spherical harmonics $Y_{\ell m}(\vec{k})$,
\beq
I(\vec{k}) = \sum\limits_{\ell= 0}^{\infty} \sum\limits_{m= -\ell}^{m= +\ell} a_{\ell m}\, Y_{\ell m}(\vec{k}).
\label{eq:multipoledecomp}
\eeq
where in the following, $I$ describes \gr{} intensities or the dimensionless maps of binned \gr{} event numbers. The  $C_{\ell}$ are defined as the covariance of the uncorrelated coefficients $a_{\ell m}$,
\begin{align}
\langle a_{\ell m} a^*_{\ell' m'}\rangle = C_{\ell}\,\delta_{\ell\ell'}\,\delta_{mm'}\quad\Rightarrow\quad C_{\ell} = \langle |a_{\ell m}|^2\rangle\,.
\label{eq:cl_covariance}
\end{align}
For a statistically isotropic field with $\langle a_{\ell m} \rangle = 0$,  the ensemble average of the $a_{\ell m}$,
\beq
\widehat{C}_\ell = \frac{1}{2\ell + 1} \sum\limits_m |a_{\ell m}|^2 \,,
\label{eq:aps}
\eeq
provides an unbiased estimator for $C_{\ell}$. With the common normalisation $\int | Y_{\ell m}(\vec{k}) |^2\,\dd\Omega = 1$ it is $[C_{\ell}] = [I]^2\times \unit{sr}$, and we will denote this dimensional \textit{intensity APS} in the following with a superscript, $C_{\ell}^I$. Throughout this paper, we assume the residual cosmic-ray background for Earth-bound \gr{} observation to be isotropic with no intrinsic  power ($C_{\ell,\,\rm bck}^I=0$). However, the shot noise from \nevents{} disjoint total events, binned in \npix{} pixels, adds a noise power $C_{\rm N}^I$ (indicated by the subscript N for \textit{noise}) to the measurement,
\beq
C_{\rm N}^I  = \frac{4\pi\,\unit{sr}}{\fsky}\;\frac{\nevents{}}{\npix{}^2}\,,
\label{eq:cP_I_events}
\eeq 
where $f_{\rm sky}=\frac{1}{4\pi\,\rm sr}\,\int W(\vec{k})\, \dd \Omega$ is the unmasked part of the sky. With this, we reconstruct the full-sky equivalent signal APS, $C_{\ell,\,\rm sig,\,\rm data}^I$, from a measured APS, $C_{\ell,\,\rm raw}^I$, to\footnote{Note that the wording ``signal'' and ``noise'' are throughout this paper not used in the usual meaning of \gr{} signal and cosmic-ray background, but to distinguish an intrinsic physical anisotropy (in both $\gamma$- and cosmic rays) against the shot noise from events binned in discrete pixels.}
\beq
C_{\ell,\,\rm sig,\,\rm data}^I= \left(C_{\ell,\,\rm full\mbox{-}sky}^I - C_{\rm N}^I\right)\times(W_{\ell}^{\rm beam})^{-2}\,,
\label{eq:signalAPS}
\eeq
where the power is unfolded by the beam suppression of a radially symmetric point spread function (PSF) $\dd P/\dd \theta$  \cite{1995PhRvD..52.4307K},
\beq
W_{\ell}^{\rm beam}(E) =  \frac{2\pi\,\rm sr}{\Omega_B}\int\limits_{-1}^1\;\mathcal{P}_{\ell}(\cos(\theta))\times \frac{\dd P}{\dd \theta}(\theta,\,E) \;\dd (\cos\theta)\,,
\label{eq:beam_function}
\eeq
and $\mathcal{P}_{\ell}$ are the Legendre polynomials of the $\ell$-th order, and  $\Omega_B=\int \dd P/\dd \theta\, \dd \Omega$. Finally, we make use of the approximation
\beq
C_{\ell,\,\rm full\mbox{-}sky}^I \approx \frac{C_{\ell,\,\rm raw}^I}{f_{\rm sky}}
\label{eq:cfullsky_approx}
\eeq
 to account for the sampling of the APS on a limited sky patch. More precisely, a masking  of $I$ with a window $W(\vec{k})$ in angular space results in a convolution in $\ell$-space \citep{2001PhRvD..64h3003W,2002ApJ...566...19K,2004MNRAS.353...43P},
\beq
C_{\ell,\,\rm raw}^I  = \frac{1}{2\ell + 1}\sum\limits_{\ell'}\sum\limits_{mm'} \,M_{\ell \ell' m m'}\,C_{\ell',\,\rm full\mbox{-}sky}^I\,,
\label{eq:pseudo_cl}
\eeq
with the convolution kernel $M_{\ell \ell'}$,
\beq
M_{\ell \ell' m m'} = \int_{\Omega} \, W(\vec{k})\, Y^*_{\ell m}(\vec{k})\, Y_{\ell' m'}(\vec{k})\,\dd \Omega\,.
\label{eq:Mllmm}
\eeq

At sufficiently large $\ell$, it is $M_{\ell \ell' m m'}^{-1} \approx \frac{2\ell +1}{\fsky}\;\delta_{\ell \ell'}\delta_{ m m'}$  and \cref{eq:cfullsky_approx} holds \citep{2002ApJ...566...19K}. However, like in Euclidean Fourier transformation, sharp window edges in angular space cause global spectral leakage artefacts in $\ell$-space. In \cref{sec:aps_analysis_background}, we will discuss these effects and the amount of spectral leakage for various CTA field of view shapes.

Due to its large residual background, CTA is unable to determine the absolute level of the DGRB intensity, $I_{\rm DGRB}$, which we are ultimately interested in probing for anisotropies. Therefore, we will work with the dimensionless \textit{fluctuation APS}, $C_{\ell}^F := C_{\ell}^I/I^2$. For event maps, the fluctuation APS is connected to the intensity APS via
\beq
C_{\ell,\,\rm sig}^F = \fsky^2\times \left(\frac{\npix}{\nevents}\right)^2 \times C_{\ell,\,\rm sig}^I\,.
\label{eq:flucAPS}
\eeq

Under the assumption that $I$ consists of two components, $\nevents = N_{\rm bck} +\nsig$,  where the background events $N_{\rm bck}$ contain zero intrinsic small-scale anisotropy (despite the shot noise power), the fluctuation APS of the DGRB component can be estimated as
 \beq
C_{\ell,\, \rm DGRB}^F = C_{\ell,\, \rm sig}^F \times \left(\frac{N_{\rm ev}}{\nsig}\right)^2\,.
\label{eq:APS_subsourceclass}
\eeq
Note that the estimated number \nsig{} of \gr{} events in the data set must be calculated with some a priori knowledge, e.g, using a spectrum measured by the \fermi{} and the expected CTA instrumental response. This is what we will later do to interpret the  sensitivity towards anisotropies in CTA data in the context of the DGRB.

We finally use a maximum likelihood (ML) approach to estimate the significance of an APS detection and to fit the signal parameters. We tailor our analysis to the assumption that the DGRB is dominated by unresolved point sources, which show an APS constant in multipole \cite{2009PhRvD..80b3520A}.\footnote{A power-law scaling of the APS within the accessible multipole range, $C_{\ell,\,\rm sig,\,model}  = C_{0}\times \left({\ell}/{\ell_0}\right)^s$ with $s\leq 0$ can be assumed to probe extended source classes; in particular for the expected APS from Galactic DM subhalos \cite{2013MNRAS.429.1529F,2013PhRvD..87l3539A,2014MNRAS.442.1151C,2015MNRAS.447..939L,2016JCAP...09..047H}. It is straightforward to correspondingly generalise the here described likelihood analysis, respecting  a couple of caveats (e.g., adopting the correct test statistic discussed in the later \cref{sec:aps_analysis_background}).} This constant power is indicated by the subscript P for \textit{Poisson power}:
\beq
C_{\ell,\,\rm sig,\,model}  = C_{\rm P}\,.
\label{eq:csig_model}
\eeq

We pursue an analysis binned in multipoles $\ell$. A binned analysis turns out be necessary to suppress remaining masking artefacts in the range of analysis, i.e., a correlation between neighbouring multipoles. We follow the findings by \cite{2016PhRvD..94l3005F} and use the unweighted arithmetic mean in each bin, $\left\langle {C_{\rm sig,\,data}}\right\rangle_{i} = \sum_{\ell_{\rm min,\,i}}^{\ell_{\rm max,\,i}}\;C_{\ell,\,\rm sig,\,data}/\Delta \ell_i$ with $\Delta \ell_i= \ell_{\rm max,\,i} - \ell_{\rm min,\,i}+1$. For an isotropic field, \cref{eq:cl_covariance}, the probability density of $C_{\ell}$ follows a $\chi^2_{2\ell +1}$ distribution and the  error on the binned  $\left\langle {C_{\rm sig,\,data}}\right\rangle_{i}$ is \cite{1995PhRvD..52.4307K}
\begin{align}
\left\langle \sigma_{\ell,\,\rm model}^{\;2}\right\rangle_{i} =\frac{2}{\,f_{\rm sky} \,(\Delta \ell_i)^2}\, \sum_{\ell_{\rm min,\,i}}^{\ell_{\rm max,\,i}}\; \frac{ \left(C_{\rm P}^I+C_{\rm N}^I\times(W_{\ell}^{\rm beam})^{-2}\right)^{2}}{2\ell +1}\,, 
\label{eq:APS_errorterm_fsky_binned}
\end{align}
where we assume $\langle \sigma_{\ell,\,\rm data}^{\;2}\rangle_{i}\approx \langle \sigma_{\ell,\,\rm model}^{\;2}\rangle_{i}\equiv\langle \sigma_{\ell}^{\;2}\rangle_{i}$. We find that a logarithmic binning with  $\sim 5$ bins  per $\ell-$decade is able to eliminate the  multipole correlation due to the masking (which breaks the assumption of \cref{eq:cl_covariance}) and to assure the correct error estimation according to \cref{eq:APS_errorterm_fsky_binned}. 

With these definitions, we construct the likelihood function 
\beq
\mathscr{L}\left(C_{\rm P}^I\,|\, \vec{C}_{\rm sig,\,data}^I\right) =  \prod\limits_{i}^{N_{\rm bins}} \frac{1}{\sqrt{2\pi\,\langle \sigma_{\ell}^{\;2}\rangle_i} }\,\exp\left(- \frac{\left[\left\langle C_{\rm sig,\,data}^I\right\rangle_i - C_{\rm P}^I\; \right]^2}{2\,\langle \sigma_{\ell}^{\;2}\rangle_{i} }\right)
\label{eq:likelihood_APS}
\eeq
where  $ \vec{C}_{\rm sig,\,data}^I$ denotes the ensemble of reconstructed measured multipoles $C_{\ell,\,\rm sig,\,data}^I$.

We then maximise the logarithm of the likelihood $\mathscr{L}$, \cref{eq:likelihood_APS}, under the constraint $C^I_{\rm P}\equiv 0$ and with $C^I_{\rm P}$ allowed to vary, and use the ratio of these maximised log-likelihoods as test statistic, 
\begin{align}
\mathrm{TS} &= -2\,\log\left(\frac{\mathscr{L}\left(C_{\rm P}^I \equiv 0\,|\, \vec{C}_{\rm sig,\,data}\right)}{\mathscr{L}\left(\widehat{C}_{\rm P}^I\,|\, \vec{C}_{\rm sig,\,data}\right)}\right) \nonumber\\[0.3cm]
 &= \sum\limits_{i}^{N_{\rm bins}}\left( \frac{\left[\left\langle C_{\rm sig,\,data}^I\right\rangle_i - \widehat{C}_{\rm P}^I \; \right]^2}{ \left\langle \widehat{\sigma}_{\ell}^{\;2}\right\rangle_{i} } - \frac{\left\langle C_{\rm sig,\,data}^I\right\rangle_i^{\;2}}{ \left\langle \widehat{\widehat{\sigma}}_{\ell}^{\;2} \right\rangle_{i} } + \log  \frac{\left\langle \widehat{\sigma}_{\ell}^{\;2}\right\rangle_{i} }{ \left\langle \widehat{\widehat{\sigma}}_{\ell}^{\;2}\right\rangle_{i} }\right)\,,
\label{eq:tsAPS}
\end{align}
to quantify the significance of some signal $C_{\rm P}^I\neq 0$ being present in the data. Here, $\widehat{C}_{\rm P}^I$ and $\langle\widehat{\sigma}_{\ell}^{\;2}\rangle_i$ are the ML estimators for the signal APS amplitude $C_{\rm P}^I$ and its variance according to \cref{eq:APS_errorterm_fsky_binned};  $\widehat{\widehat{\sigma}}_{\ell}^{\;2}$ denotes the estimated variance under the constraint $C^I_{\rm P} \equiv 0$ of no signal present in the data.


\section{CTA instrumental characteristic and large-area survey model}
\label{sec:cta}

\subsection{Instrumental performance of the southern CTA}
\label{sec:cta_instrumentcharact}
The Cherenkov Telescope Array (CTA) will be the next-generation Imaging Atmospheric Cherenkov Telescope (IACT) array, consisting of two separate arrays to be erected at the Paranal site (Chile) in the southern hemisphere  and on La Palma (Spain) in the northern hemisphere. In this paper, we base the  performance of CTA on the recently published \texttt{prod3b-v1} instrumental performance simulation\footnote{\url{http://www.cta-observatory.org/science/cta-performance}} and solely on the southern array, comprising 4 Large-Size Telescopes (LSTs), 25 Mid-Size Telescopes (MSTs), and 70 Small-Size Telescopes (SSTs). We rely on the southern CTA for the sake of simplicity and to consider the presence of SSTs (which are planned to be part only of the southern array), yielding a higher sensitivity in the TeV regime. In contrast to the earlier \texttt{prod2} characteristics, the \texttt{prod3b} performance estimation provides reliable results for Cherenkov light  hitting the camera planes under an incidence angle $\vartheta$ \textit{off-axis} of the telescopes'  pointing direction, crucial for the study of the CTA survey performance. Note that little has changed with respect to the projected \textit{on-axis} CTA performance ($\vartheta=0$) between \texttt{prod2} (as used, e.g., in our previous study \cite{2016JCAP...09..047H}) and \texttt{prod3b}. 

The instrumental performance characteristics are the result of extensive Monte-Carlo (MC) simulations. These simulations comprise the modelling of all steps in the Cherenkov light emission and detection, from the shower development in the atmosphere to the response of the photosensors in the cameras. Because of the high computational costs, calculations are done in relatively coarse bins of the parameters (elevation and azimuth angle of the pointing, \gr{} energy, off-axis angle,\ldots) such that we apply suitable analytical descriptions of the various instrument response quantities fitted to their discrete tabulated values. Such a fitting approach instead of  interpolation is necessary for this study, where we intend to avoid artefacts in the APS which result from the numerics of the simulation (and would not be present in a CTA data set). We assume that all observations are taken at a zenith angle of $20\degs$, with equal proportions of southern and northern pointings. Besides the physics in the atmosphere and the instrumental layout, the performance characteristics depend on the chosen analysis cuts. If not explicitly stated differently, all event simulations are based on the CTA performance according to an event selection optimised for $\unit[5]{h}$ of observation and without spatial cuts, other than removing events at $\vartheta>6\degs$ from the camera centre. 

\begin{figure}[t]
\centering
\includegraphics[width=0.62\textwidth]{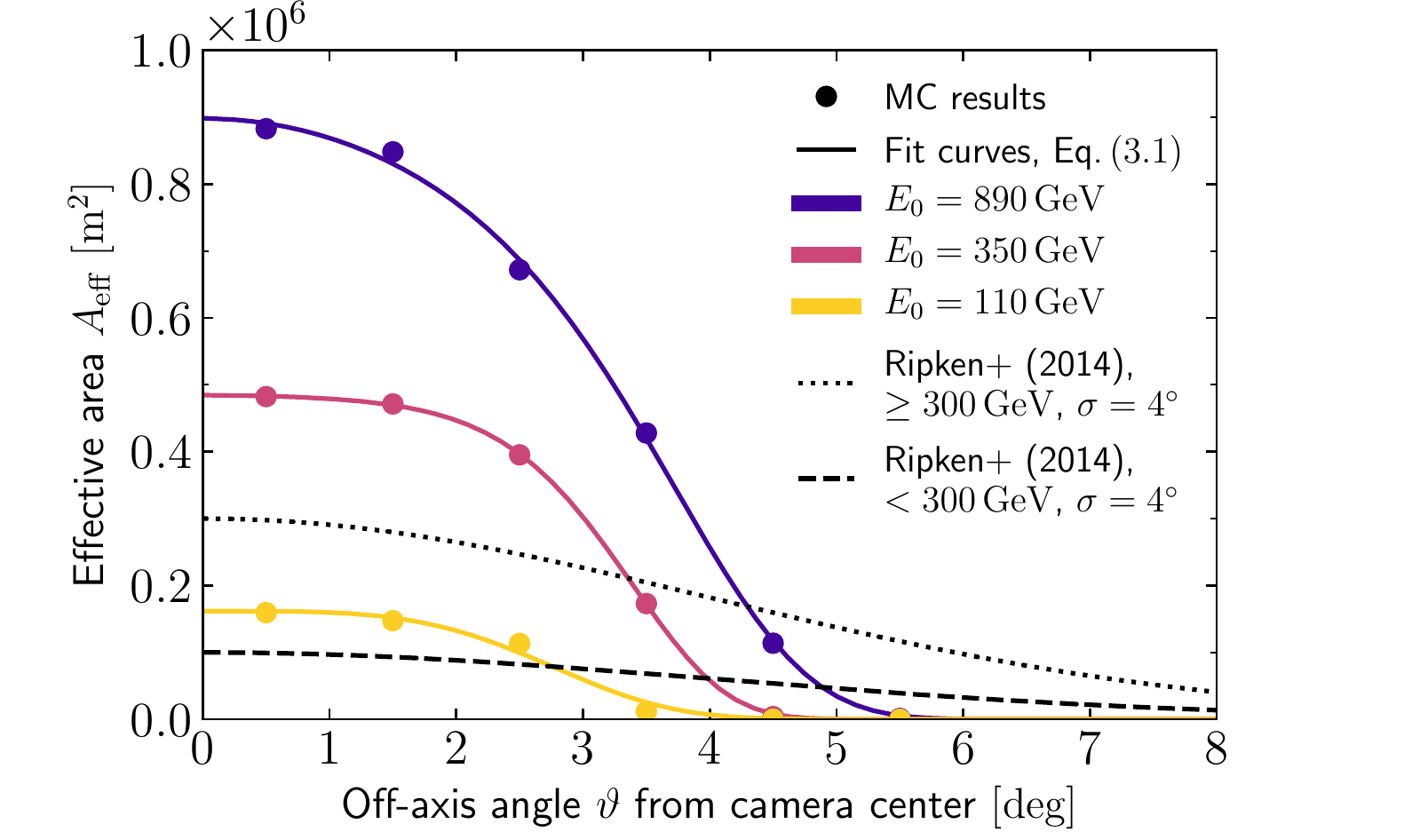}
\caption{
Effective areas for the \texttt{prod3b} southern CTA off-axis performance at different reconstructed \gr{} energies $\lesssim \unit[1]{TeV}$. The dashed and dotted curves additionally show the earlier CTA effective area models from \cite{2014JCAP...01..049R}.
}
\label{fig:CTA_EffAreaOffAxis}
\end{figure}

\begin{figure}[t]
\centering
    \includegraphics[width=0.62\textwidth]{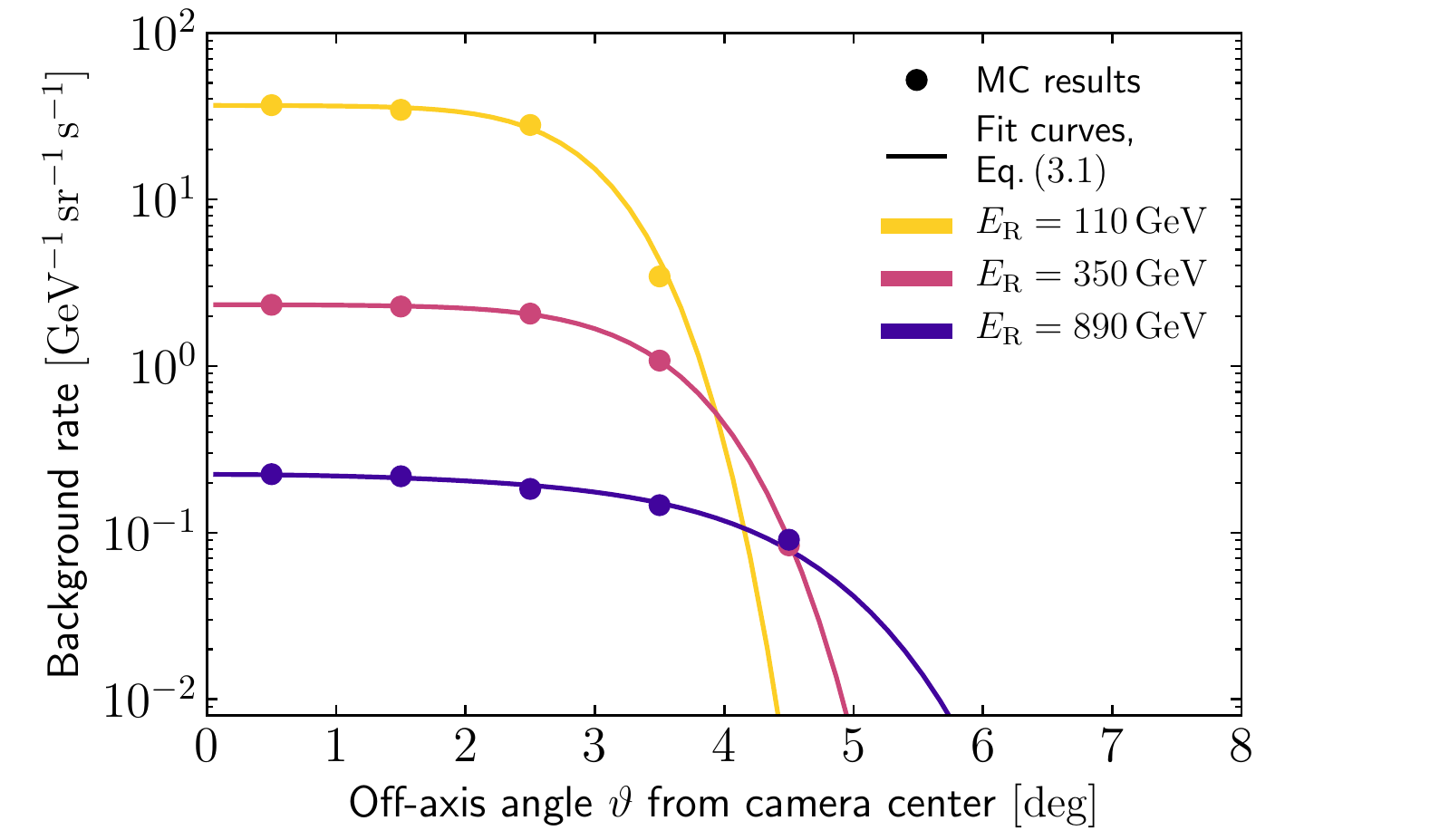}
  \captionof{figure}[Background rates for the CTA off-axis performance]{
Background rates for the \texttt{prod3b} southern CTA off-axis performance at different reconstructed \gr{} energies $\lesssim \unit[1]{TeV}$.
}
 \label{fig:CTA_BckRateOffAxis}
\end{figure}

We describe the dependence of the effective \gr{} collection area, $A_{\rm eff}$, and the cosmic-ray residual background rate on the angle $\vartheta$, using a higher-order Gaussian function:
\beq
f(\vartheta) = A\, \exp\left[-\frac{1}{2}\,\left(\left( \frac{\vartheta}{B} \right)^6 + \left( \frac{\vartheta}{C} \right)^4 + \left( \frac{\vartheta}{D} \right)^2 \right)\right]\,.
\label{eq:fitIRF}
\eeq
While previous studies, e.g., \cite{2013APh....43..317D,2014JCAP...01..049R}, have modelled the off-axis behaviour of the CTA effective area and background rate with a Gaussian function ($B=C\equiv\infty$), the modelling by \cref{eq:fitIRF} takes into account the plateau shape of the responses around the pointing position. \Cref{fig:CTA_EffAreaOffAxis} shows the off-axis dependence of $A_{\rm eff}$ for  the southern CTA, comparing the MC results for six bins between $\vartheta=0\degs$ and $\vartheta=6\degs$ with the fit by \cref{eq:fitIRF}. This clearly reveals a deviation from a Gaussian shape, in particular at the lowest energies. \Cref{fig:CTA_EffAreaOffAxis} also displays the Gaussian CTA effective area model in the earlier study  on APS measurements with IACTs \cite{2014JCAP...01..049R}. \Cref{fig:CTA_BckRateOffAxis} shows our off-axis fitting approach for some exemplary \texttt{prod3b} residual background rates (with the abscissa in log-scale, such that a Gaussian scaling would be represented by a parabola). When speaking about the CTA field of view (FOV) in the following, we refer to the energy-dependent off-axis shape of the residual background rate. We define the FOV radius, $\vartheta_{\rm fov}$, via
\beq
\int_{S^2}\frac{\dd N_{\rm bck}}{\dd \Omega\, \dd t}(\vartheta)\,\dd\Omega = \frac{\dd N_{\rm bck}}{\dd \Omega \,\dd t}(\vartheta=0)\times \left(1 - \cos\vartheta_{\rm fov} \right)\times \unit[2\pi]{sr}\,.
\label{eq:fov-radius}
\eeq

In \cref{tab:off-axis-performance-prod3b}, we show the on- and off-axis event rates above various energy thresholds for the CTA instrumental response used throughout this paper.\footnote{Note that in \cite{2016JCAP...09..047H}, we have presented a similar table (Table 2) for the {\tt prod2} on-axis rates.} It can be seen that the FOV radius according to \cref{eq:fov-radius} increases with energy, which causes the background over the full FOV to decrease more slowly with energy than the on-axis background rate alone. We also display the expected event rate from a DGRB spectrum after \cite{2015ApJ...799...86A}, using their foreground model B, which leaves the largest fraction of unassociated \grs{} to the DGRB:
\beq
I_{\rm DGRB} =  1.12\times 10^{-7}\times 
\left(\frac{E}{\unit[100]{MeV}}\right)^{-2.28}\times \exp\left(-\frac{E}{\unit[206]{GeV}}\right)\;\unit{cm^{-2}\,s^{-1}\,sr^{-1}\,MeV^{-1}}\,.
\label{eq:dgrb_ackermann2015}
\eeq

In \cref{tab:off-axis-performance-ripken}, we show the CTA instrumental performance according to the previous work from \cite{2014JCAP...01..049R}, to be compared with our model in \cref{tab:off-axis-performance-prod3b}. As shown in \cref{fig:CTA_EffAreaOffAxis}, the authors of \cite{2014JCAP...01..049R} assumed a much wider FOV (both for the effective areas and background rates), however, with smaller on-axis effective areas. Combined, they obtain rather similar \gr{} event rates compared to our {\tt prod3b} model. Also, their assumption of background rates between $10-\unit[100]{Hz}$ above $\unit[100]{GeV}$ has remained valid. However, we will argue in \cref{sec:aps_analysis_background} that the non-Gaussian and, compared to \cite{2014JCAP...01..049R}, substantially smaller FOV in the {\tt prod3b} model provides a major obstacle for analysing deep-field observations for \gr{} anisotropies.

\begin{table}[t]
 \centering
 \resizebox{\textwidth}{!}{%
  \begin{tabular}{|c|c|cc|cc|cc|} \hline
   & &\multicolumn{2}{c |}{DGRB event rate} & \multicolumn{2}{c |}{Background rate} & \multicolumn{2}{c |}{{\grs}/background} \\
\multirow{-2}{*}{Energy} & \multirow{-2}{*}{FOV radius} & on-axis & total FOV  &  on-axis & total FOV &  \multicolumn{2}{c |}{ratio}  \\
\multirow{-2}{*}{threshold}& \multirow{-2}{*}{$\vartheta_{\rm fov}$} & $\mathrm{[Hz\;\deg^{-2}]}$&  $\mathrm{[Hz]}$&  $\mathrm{[Hz\;\deg^{-2}]}$& $\mathrm{[Hz]}$&  on-axis & total FOV  \\\hline
$30\,\mathrm{GeV}$  & $2.4\degs$ &$8.8\times 10^{-4}$ & $1.6\times 10^{-2}$ & 3.6   & 66.2 & $2.4\times 10^{-4}$ & $2.5\times 10^{-4}$\\
$100\,\mathrm{GeV}$ & $3.3\degs$ &$3.6\times 10^{-4}$ & $1.0\times 10^{-2}$ & 1.1   & 37.0  & $3.3\times 10^{-4}$ & $2.7\times 10^{-4}$\\
$300\,\mathrm{GeV}$ & $4.1\degs$ &$4.6\times 10^{-5}$ & $1.6\times 10^{-3}$ & 0.24  & 12.4  & $2.0\times 10^{-4}$ & $1.3\times 10^{-4}$\\
$1\,\mathrm{TeV}$ & $5.3\degs$ &$3.2\times 10^{-7}$ & $1.3\times 10^{-5}$ & 0.056  & 5.0  & $5.8\times 10^{-6}$ & $2.5\times 10^{-6}$\\\hline
  \end{tabular}
  }
 \caption[Diffuse {\gr} and background rates  for the southern CTA]{
DGRB event rates and background rates (without dead time correction), integrated over energies above different lower thresholds up to $E_{\rm max}=\unit[100]{TeV}$. The DGRB rates correspond to the DGRB intensity spectrum according to \cref{eq:dgrb_ackermann2015}. Events in the ``total FOV'' comprise all events within the angular direction $\vartheta_{\rm cut}=6\degs$ from the camera pointing position.
  }
\label{tab:off-axis-performance-prod3b}
\end{table}

\begin{table}[t]
 \centering
 \resizebox{\textwidth}{!}{%
  \begin{tabular}{|c|c|cc|cc|cc|} \hline
   & &\multicolumn{2}{c |}{DGRB event rate} & \multicolumn{2}{c |}{Background rate} & \multicolumn{2}{c |}{{\grs}/background} \\
\multirow{-2}{*}{Energy} &  \multirow{-2}{*}{FOV radius} & on-axis & total FOV  &  on-axis & total FOV &  \multicolumn{2}{c |}{ratio}  \\
\multirow{-2}{*}{threshold}& \multirow{-2}{*}{$\vartheta_{\rm fov}$} & $\mathrm{[Hz\;\deg^{-2}]}$&  $\mathrm{[Hz]}$&  $\mathrm{[Hz\;\deg^{-2}]}$& $\mathrm{[Hz]}$&  on-axis & total FOV  \\\hline
$100\,\mathrm{GeV}$ & $5.7\degs$ &$1.6\times 10^{-4}$ & $1.6\times 10^{-2}$ & 0.99   & 100  & $1.6\times 10^{-4}$ & $1.6\times 10^{-4}$\\
$300\,\mathrm{GeV}$ & $5.7\degs$ &$2.6\times 10^{-5}$ & $2.6\times 10^{-3}$ & 0.099  & 10  & $2.6\times 10^{-4}$ & $2.6\times 10^{-4}$\\\hline
  \end{tabular}
  }
 \caption[Diffuse {\gr} and background rates  for the southern CTA]{
Model of the CTA performance from \cite{2014JCAP...01..049R}, according to their tables 1 \& 2, $\sigma_{\rm fov}=4\degs$, and the assumption of 100/10 Hz background rates above 100/300 GeV. Again, the DGRB rates are computed for the model B from \cite{2015ApJ...799...86A}. Note that for a Gaussian off-axis acceptance, $\vartheta_{\rm fov}\approx\sqrt{2}\,\sigma_{\rm fov}$.
  }
\label{tab:off-axis-performance-ripken}
\end{table}

The finite CTA angular resolution causes a suppression of the APS at angular scales below the resolution, expressed by \cref{eq:beam_function}.  The CTA PSF worsens at lower \gr{} energies $E$ and for large angles $\vartheta$ of the incident \gr{} in the camera field, i.e. it is $\dd P/\dd \theta=\dd P/\dd \theta(\theta;\,E,\,\vartheta)$. We model $\dd P/\dd \theta(E,\,\theta,\,\vartheta)$  as a two parameter King function \cite{2016A&A...593A...1K} 
\beq
 \dd P/\dd \theta(\theta;\,E,\,\,\vartheta) = \frac{1}{2\pi\sigma_{\rm king}}\,\left(1-\frac{1}{\gamma}\right)\, \left(1+\frac{1}{2\gamma}\frac{\theta^2}{\sigma_{\rm king}^2}\right)^{-\gamma}\,,
 \label{eq:fitPSF}
\eeq
owing to the fact that CTA will exhibit a  PSF with non-Gaussian tails (see the later \cref{fig:CTA_Wbeam_less}).  The two parameters $\sigma_{\rm king}=\sigma_{\rm king}(E,\,\,\vartheta)$ and $\gamma=\gamma(E,\,\,\vartheta)$ are again fitted to the MC simulations of the instrumental performance.

\subsection{A model  of the CTA extragalactic survey}
\label{sec:cta_survey}

With its large FOV, CTA will be the first IACT array to perform a large-area sky survey \cite{2017arXiv170907997C}. Within the CTA \textit{extragalactic survey key science project} \cite{2017arXiv170907997C}, it is planned to observe  $25\%$ of the sky outside the Galactic plane within the first decade of observation. For a total observation time of about $\unit[1000]{h}$ with both arrays, such a survey is projected to reach an average sensitivity to fluxes greater than $\unit[2.5\times 10^{-12}]{cm^{-2}\,s^{-1}}$ ($0.6\%$ the flux of the Crab Nebula) above $\unit[125]{GeV}$ for point sources with a spectrum similar to the one of the Crab Nebula \cite{2017arXiv170907997C}. In this paper, we adopt the same survey field as done in \cite{2016JCAP...09..047H} for the extragalactic survey, namely, a circular region around the Galactic south pole, i.e., $b < -30\degs$. As in \cite{2016JCAP...09..047H}, we assume that the whole area is covered by the southern array in $\unit[500]{h}$, while \cite{2017arXiv170907997C} propose to raster $60\%$ of the survey field in $\unit[400]{h}$ with the southern array, and the remaining area in $\unit[600]{h}$ with the northern CTA. In this work, we consider a system dead time of 5\%, reducing our effective total observation time to $T_{\rm obs} = \unit[475]{h}$. Note that this choice is fairly conservative, and CTA is targeted to reach system dead times smaller than $2\%$. On the other hand, we do not consider a loss of observation time due to the telescope slewing  between each pointing. 

\begin{figure}[t]
\centering
    \subfigure[$\Delta_{\rm fov}\approx 4\degs$ ($\nside=16$)]{
    \includegraphics[width=0.62\textwidth]{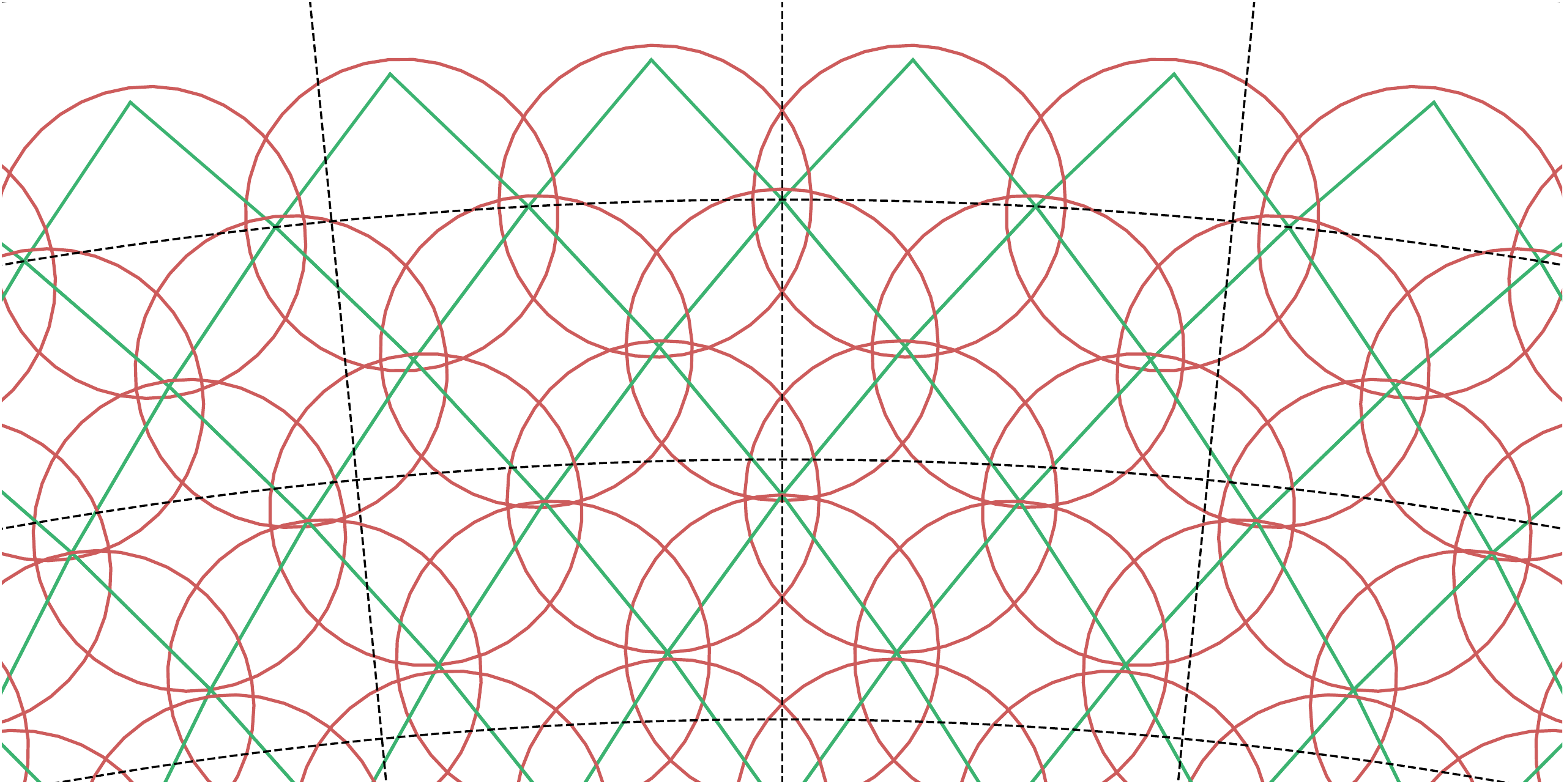}
        \fontsize{9.5pt}{11pt}\selectfont
        \begin{picture}(1,1)%
\put(1,88){\color[rgb]{0,0,0}\makebox(0,0)[l]{\smash{$b=-30\degs$}}}%
\put(1,42){\color[rgb]{0,0,0}\makebox(0,0)[l]{\smash{$b=-35\degs$}}}%
\put(1,-2){\color[rgb]{0,0,0}\makebox(0,0)[l]{\smash{$b=-40\degs$}}}%
\put(-137,140){\color[rgb]{0,0,0}\makebox(0,0)[c]{\smash{$l=0\degs$}}}%
\put(-221,140){\color[rgb]{0,0,0}\makebox(0,0)[c]{\smash{$l=+10\degs$}}}%
\put(-53,140){\color[rgb]{0,0,0}\makebox(0,0)[c]{\smash{$l=-10\degs$}}}%
\end{picture}
    }
    \subfigure[$\Delta_{\rm fov}\approx 1\degs$ ($\nside=64$)]{\includegraphics[width=0.62\textwidth]{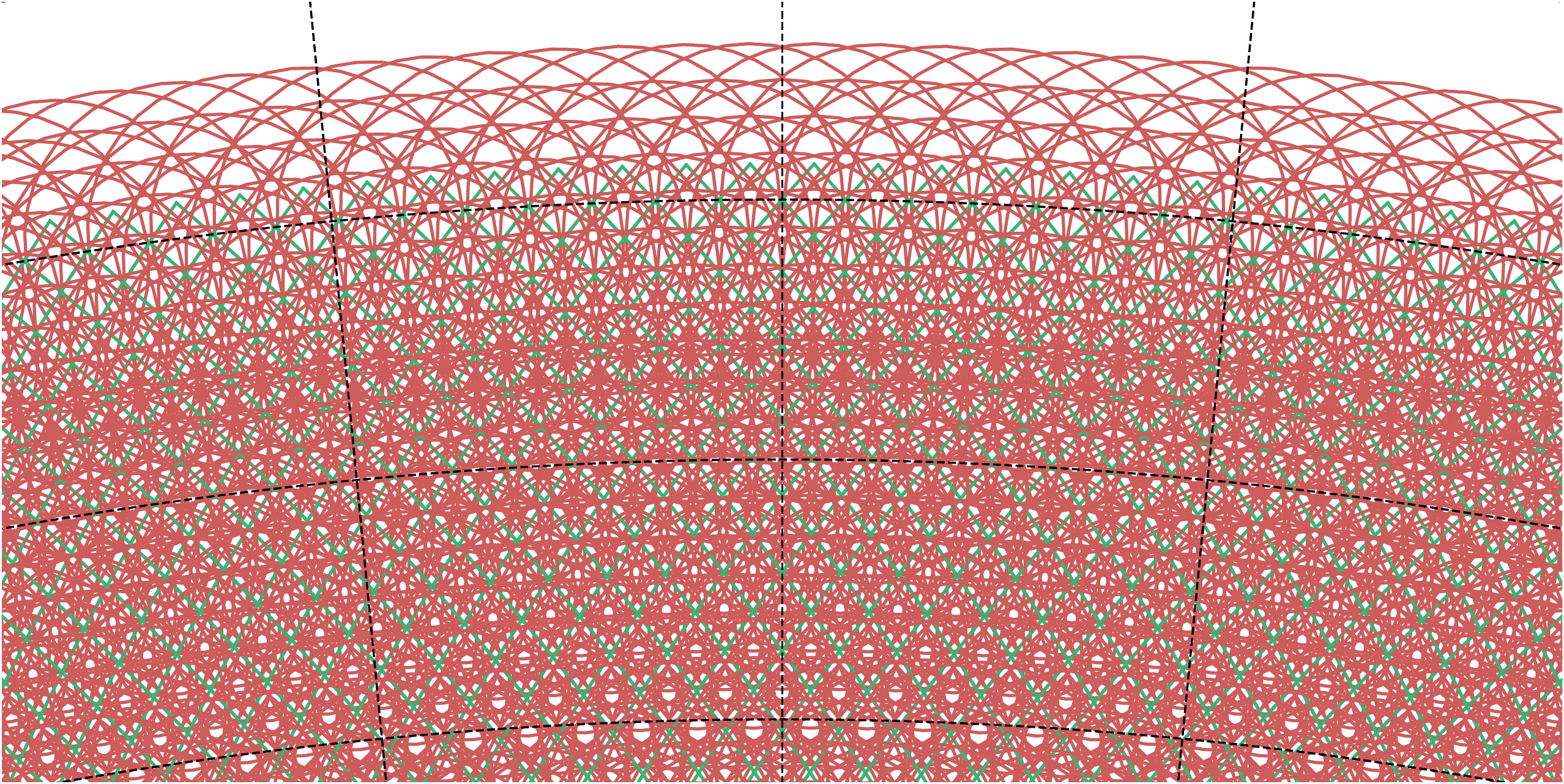}
            \fontsize{9.5pt}{11pt}\selectfont
        \begin{picture}(1,1)%
\put(1,88){\color[rgb]{0,0,0}\makebox(0,0)[l]{\smash{$b=-30\degs$}}}%
\put(1,42){\color[rgb]{0,0,0}\makebox(0,0)[l]{\smash{$b=-35\degs$}}}%
\put(1,-2){\color[rgb]{0,0,0}\makebox(0,0)[l]{\smash{$b=-40\degs$}}}%
\end{picture}
    }
\caption{{
Large-area survey tiling strategy on $\fsky=0.25$ in the \healpix{} scheme as adopted in this work. For display purpose only, we show a spacing of $\Delta_{\rm fov}\approx 4\degs$ ($\nside=16$) in the upper panel. In our default survey setup, we use $\Delta_{\rm fov}\approx 1\degs$ (lower panel) and compare to $\Delta_{\rm fov}\approx 2\degs$. The green  quadrilaterals  show the \healpix{} pixels in which the individual telescope pointings are centred. The red circles have a diameter of $6\degs$ and show the CTA field of view diameter at about $\unit[100]{GeV}$. Note that all \healpix{} pixels cover the same solid angle, however, they are not congruent.
}
\label{fig:Pointings_4deg}}
\end{figure}

The FOV size of the instrument is critical for the average survey exposure on each spot on the sky and the homogeneity of the exposure. The exposure homogeneity is in particular dependent on the pattern of the pointing distances of the individual survey observations. For the pointing pattern, we adopt a grid relying on the \healpix{} pixelisation scheme \cite{2005ApJ...622..759G}. The \healpix{}  tessellation  facilitates  equally spaced pointing positions on the sphere, where the sphere curvature becomes significant for an area as large as $\fsky=0.25$. This large-area survey telescope spacing strategy is illustrated  in \cref{fig:Pointings_4deg}. In the remainder of this work, we consider two different grid spacings: A \healpix{} grid with $\nside =32$ results in an average distance between the telescope pointings of $\Delta_{\rm fov}=1.83\degs\approx 2\degs$.\footnote{$\Delta_{\rm fov}$ is defined  as the square root of the ``pointing pixel'' area with size $\Delta_{\rm fov}^2$.} Hereby, $\fsky=0.25$ is covered by 3136 individual pointings with an observation time of  $\unit[9.6]{min}$ each. This results in an average on-axis equivalent exposure of $t_{\rm obs}\approx \unit[100]{min}$ above $\unit[100]{GeV}$ with a homogeneity of $\Delta t /t \lesssim 3\%$. In contrast,  a \healpix{} grid with $\nside =64$ is considered, yielding $\Delta_{\rm fov}=0.91\degs\approx 1\degs$ with 12,416 single observations over $\unit[145]{s}$ each,  and the same average exposure $t_{\rm obs}$ above $\unit[100]{GeV}$ with $\Delta t /t \lesssim 1\%$. 

Because of the lower sensitivity of the northern CTA above \unit[100]{GeV} compared to the southern array,  our simplified large-scale survey setup finally serves as a fairly realistic description of what can be achieved with CTA according to \cite{2017arXiv170907997C}: Using the \texttt{cssens} ctool \cite{2016A&A...593A...1K}, we find for a dead time corrected on-axis observation with $t_{\rm obs} = \unit[95]{min}$ and our selected CTA instrument response a survey sensitivity of $0.4\%$ of the Crab nebula flux above $\unit[125]{GeV}$, not more than $30\%$ better than envisaged by \cite{2017arXiv170907997C}.\footnote{Requiring a test statistic of $\rm TS = 25$ and without applying trials corrections.}

\begin{figure}[t]
\centering
\includegraphics[width=\textwidth]{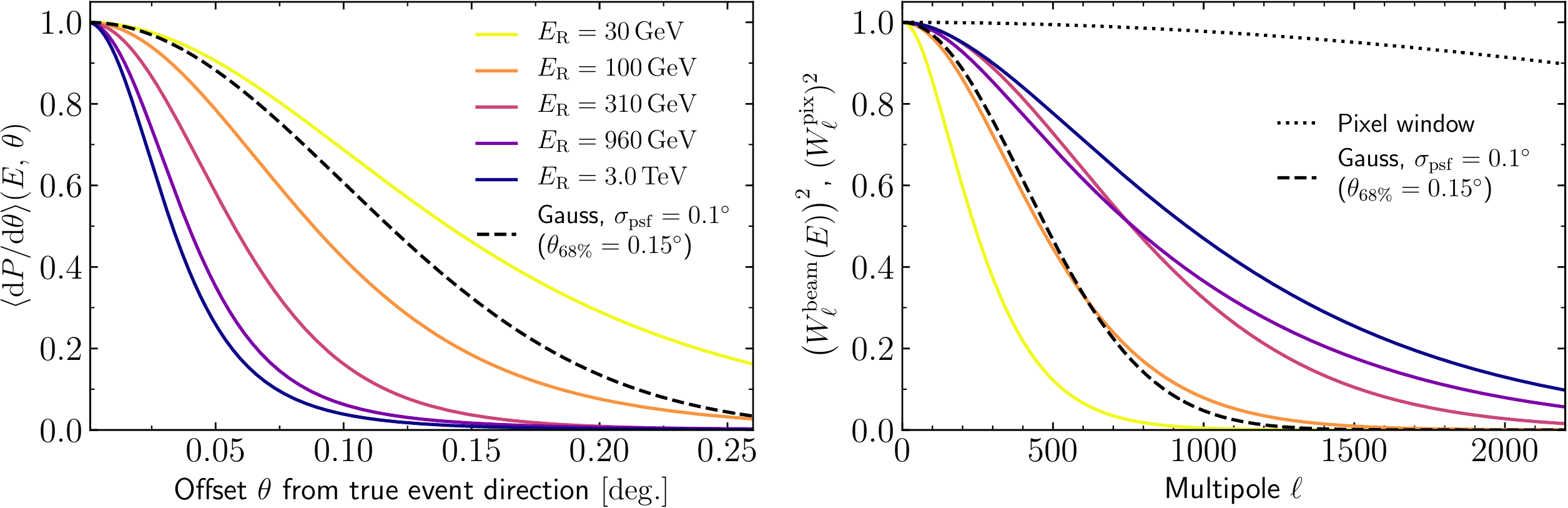}
\caption{{
\textit{Left panel:} Southern CTA angular resolution in terms of the PSF $\langle \mathrm{d}P/\mathrm{d}\theta\rangle(E,\,\theta)$ averaged over $A_{\rm eff}(E,\,\vartheta)$, in angular space at different energies. \textit{Right panel:} Corresponding window functions in multipole space. A Gaussian curve and its multipole transformation are given for comparison (dashed lines). Also, the pixel window function of a finite \healpix{} grid with resolution $\nside=2048$ is shown (dotted line on the right).
}
\label{fig:CTA_Wbeam_less}}
\end{figure}

As the angular resolution degrades with offset from the camera centre, we also have to consider an average angular resolution whose homogeneity is, like the exposure, affected by the observation pattern. We approximate an average PSF under the assumption that each spot on the sky is equally observed under all incidence angles $\vartheta$ from the camera centre up to $\vartheta_{\rm cut}=6\degs$. In a finite energy interval $[E_{\rm min},\,E_{\rm max}]$, the effective PSF additionally scales with the energy distribution of events from the \gr{} intensity spectrum $I$. Altogether, we model the effective CTA survey  PSF in a finite energy interval as:
\begin{align}
 \left\langle \frac{\dd P}{\dd \theta} \right\rangle (E,\,\theta) = \frac{\int\limits_{\vartheta_{\rm cut}} \frac{\dd P}{\dd \theta}(E,\,\theta,\,\vartheta)\times A_{\rm eff}(E,\,\vartheta)\,\dd \Omega}{\int\limits_{\vartheta_{\rm cut}}A_{\rm eff}(E,\,\vartheta)\,\dd \Omega}\,,\;  \left\langle \frac{\dd P}{\dd \theta} \right\rangle (\theta) = \frac{\int\limits_{E_{\rm min}}^{E_{\rm max}} \left\langle \frac{\dd P}{\dd \theta} \right\rangle (E,\,\theta) \times I(E)\,\dd E}{\int\limits_{E_{\rm min}}^{E_{\rm max}}I(E)\,\dd E}.
 \label{eq:avg_psf}
\end{align}
The \gr{} spectrum $I$ is chosen according to the  source class hypothesised in the analysis.   \Cref{fig:CTA_Wbeam_less} shows the averaged PSFs, $ \langle {\dd P}/{\dd \theta} \rangle (E,\,\theta) $, and their corresponding APS attenuation factors $W_{\ell}^{\rm beam}(E)$ at different energies. 

The PSF modelling determines the highest multipole $\ell_{\rm max}=1024$ up to which we perform our analysis to avoid the impact of an increasing systematic uncertainty about the PSF at larger $\ell$. This choice of $\ell_{\rm max}$ together with our chosen map resolution also allows us to ignore the power suppression of the data due to the finite bin size (``pixel window'', see right panel of \cref{fig:CTA_Wbeam_less}). Finally, we do not consider the finite energy resolution and energy bias of the CTA instrument in this study. As we investigate the CTA performance in only very coarse energy intervals, the energy resolution of the instrument is negligible.

\subsection{APS characteristic of the cosmic-ray background in CTA observations}
\label{sec:aps_analysis_background}

As shown in \cref{tab:off-axis-performance-prod3b}, CTA suffers from a large irreducible cosmic-ray background. This background constitutes the dominant challenge of power-spectral methods for Earth-bound \gr{} detectors, in contrast to the similar analyses already performed with \fermi{} data.

\begin{figure}[t]
\begin{center}
\includegraphics[width=0.62\textwidth]{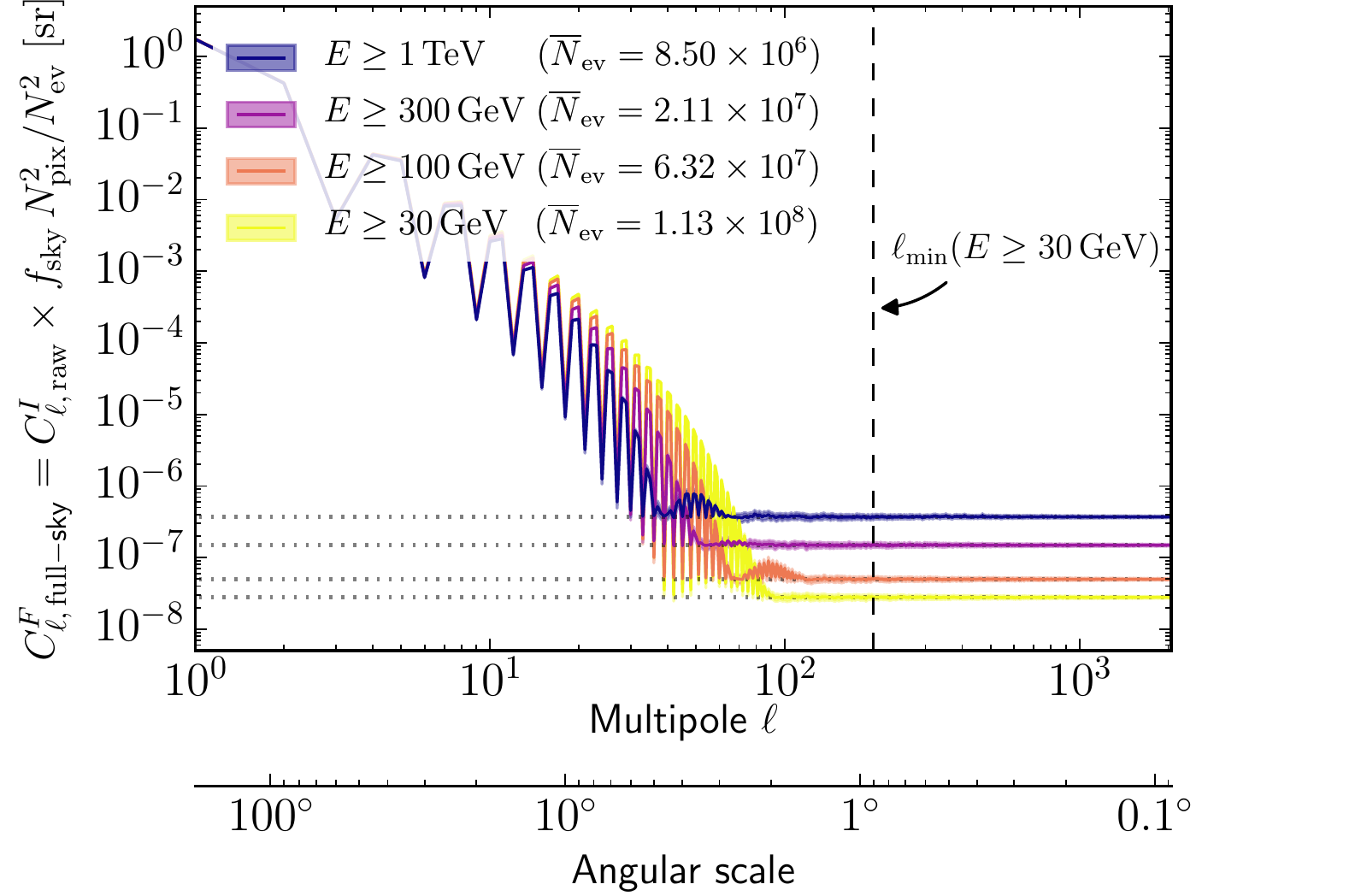}
\caption{
APS of the events from an intrinsically isotropic background in our CTA survey field on $\fsky=0.25$ and for an observation spacing of $\Delta_{\rm fov}=1\degs$. The APS is shown for four different energy thresholds. The dashed vertical line indicates one example of our lower cuts in $\ell$-space. The figure shows the mean $C_{\ell}^F$ and the 68\% 
credible intervals (shaded bands) based on 50 simulations each. The grey-dotted lines mark the estimator $C_{\rm N}^F = 4\pi\,\unit{sr} \fsky/\overline{N}_{\rm ev}$ for each energy threshold.
}
\label{fig:APS_background_4in1}
\end{center}
\end{figure}

To study the APS characteristic of this irreducible background, we use the {\tt gammalib} and {\tt ctools} frameworks \cite{2016A&A...593A...1K}\footnote{\url{http://cta.irap.omp.eu/gammalib/} and \url{http://cta.irap.omp.eu/ctools/}} to simulate hundreds of samples of the background events in our CTA extragalactic survey setup with $\Delta_{\rm fov} = 1\degs$, based on the background rates as presented in \cref{sec:cta_instrumentcharact}. Subsequently, we collect the sampled events in maps of $\npix=12\times2048^2$ spatial bins in the \healpix{} scheme,  and compute the APS of the binned event maps. \Cref{fig:APS_background_4in1} presents the resulting fluctuation APS above different energy thresholds. The shot noise estimator, \cref{eq:cP_I_events}, is indicated by the dashed horizontal lines in \cref{fig:APS_background_4in1}. It can be clearly seen that the window of the quarter-sky survey field dominates the multipole range at $\ell\lesssim 100$. According to \cref{sec:aps_analysis}, we assume (i) the approximation $C_{\ell,\,\rm full\mbox{-}sky}^I \approx {C_{\ell,\,\rm raw}^I}/{f_{\rm sky}}$ to be valid in a regime unaffected by the window above some $\ell_{\rm min}$, (ii) the same for the error described by \cref{eq:APS_errorterm_fsky_binned}, and (iii) a signal hypothesis with only one additional free parameter in the maximisation of the likelihood \cref{eq:likelihood_APS}, namely, $C_{\rm P}\geq 0$. If all these conditions are satisfied, the distribution of TS values from many samples of our background maps must follow a $\chi^2$ distribution with one degree of freedom \cite{2011EPJC...71.1554C},
\beq
p(\rm{TS}) = \frac{1}{2}\, \delta(\rm{TS}) +  \frac{1}{2}\, \chi^2_{k=1}(\rm{TS})\;.
\label{eq:ts_distribution}
\eeq
Requiring the reproduction of this statistic, we determine
\begin{align}
\ell_{\rm min}(\unit[30]{GeV}\leq E_\gamma < \unit[100]{GeV}) &= 200\,,\nonumber\\
\ell_{\rm min}(\unit[100]{GeV}\leq E_\gamma < \unit[300]{GeV}) &= 150\,,\nonumber\\
\ell_{\rm min}(\unit[300]{GeV}\leq E_\gamma <\unit[1]{TeV}) &= 120,\,\nonumber\\
\ell_{\rm min}(E_\gamma \geq \unit[1]{TeV}) &= 200 \label{eq:lmin}
\end{align}
as lower  limits in $\ell$-space for our analysis.\footnote{For the integrated sensitivity, we use the most conservative $\ell_{\rm min}=200$.} As mentioned in \cref{sec:aps_analysis}, we find that a binning of the signal and its error in multipoles is crucial to reproduce \cref{eq:ts_distribution}. This procedure worked well for all energy up to $E \leq \unit[1]{TeV}$, whereas at higher energies, window artefacts can no longer be removed by binning for $\ell\lesssim 1000$. For these highest energies, we use $\rm{TS}_{95\%,\,E \geq \unit[1]{TeV}} = 8.0$ empirically obtained from our MC simulations, but we emphasise that the results for this bin should be treated with some caution. The cuts $\ell_{\rm min}$ finally define the angular scales $\alpha_{\rm max}\approx 180\degs/\ell_{\rm min}\leq1.5\degs$ below which CTA will be able to probe  anisotropies.

\begin{figure}[t]
\centering
\includegraphics[width=\textwidth]{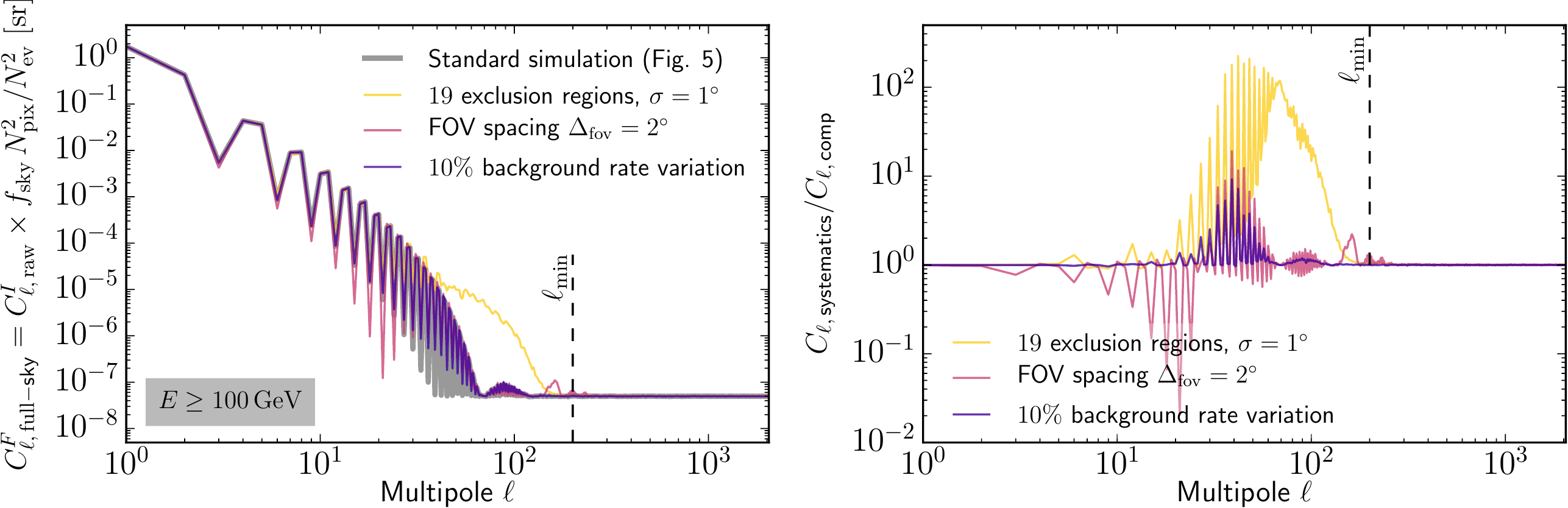}
\caption{{
The impact of various deviations from the standard survey simulation onto the residual background APS of events with $E\geq\unit[100]{GeV}$. The relative change of power at each multipole $\ell$ compared to the spectrum presented in \cref{fig:APS_background_4in1} is shown on the right.
}
\label{fig:APS_background_systematics_combined}
}
\end{figure}

Possible deviations of the available data set from the standard survey setup considered in the previous paragraph alter the background APS. Therefore, we have studied (i) the presence of 19 circular exclusion regions with a Gaussian mask and $\sigma=1\degs$ (corresponding to the 19 up-to-date known VHE \gr{} sources in the chosen survey field), (ii) a coarser survey grid with $\Delta_{\rm fov}=2\degs$, and (iii) a 10\% variation of the background rate between the individual observation pointings in the survey  due to different weather conditions and calibration uncertainty. The latter is a rather conservative estimate, as CTA is required to provide a much more stable data rate over time on all timescales. \Cref{fig:APS_background_systematics_combined} compares the absolute (left) and relative (right) APS of these varied setups to our standard case from \cref{fig:APS_background_4in1}. The dashed vertical lines in these panels show the $\ell_{\min}$ satisfying the background hypothesis $C_{\ell} = C_{\rm N}$ in our standard survey setup. We conclude that neither the exclusion of some dozens of regions in the survey nor a varying event rate significantly pollute the APS above the threshold $\ell_{\rm min}$ according to  \cref{eq:lmin}. However, a coarser survey pattern with $\Delta_{\rm fov} > 1\degs$ in fact adds significant contamination to multipoles above $\ell \geq 120$.

\begin{figure}[t]
\begin{center}
\includegraphics[width=0.62\textwidth]{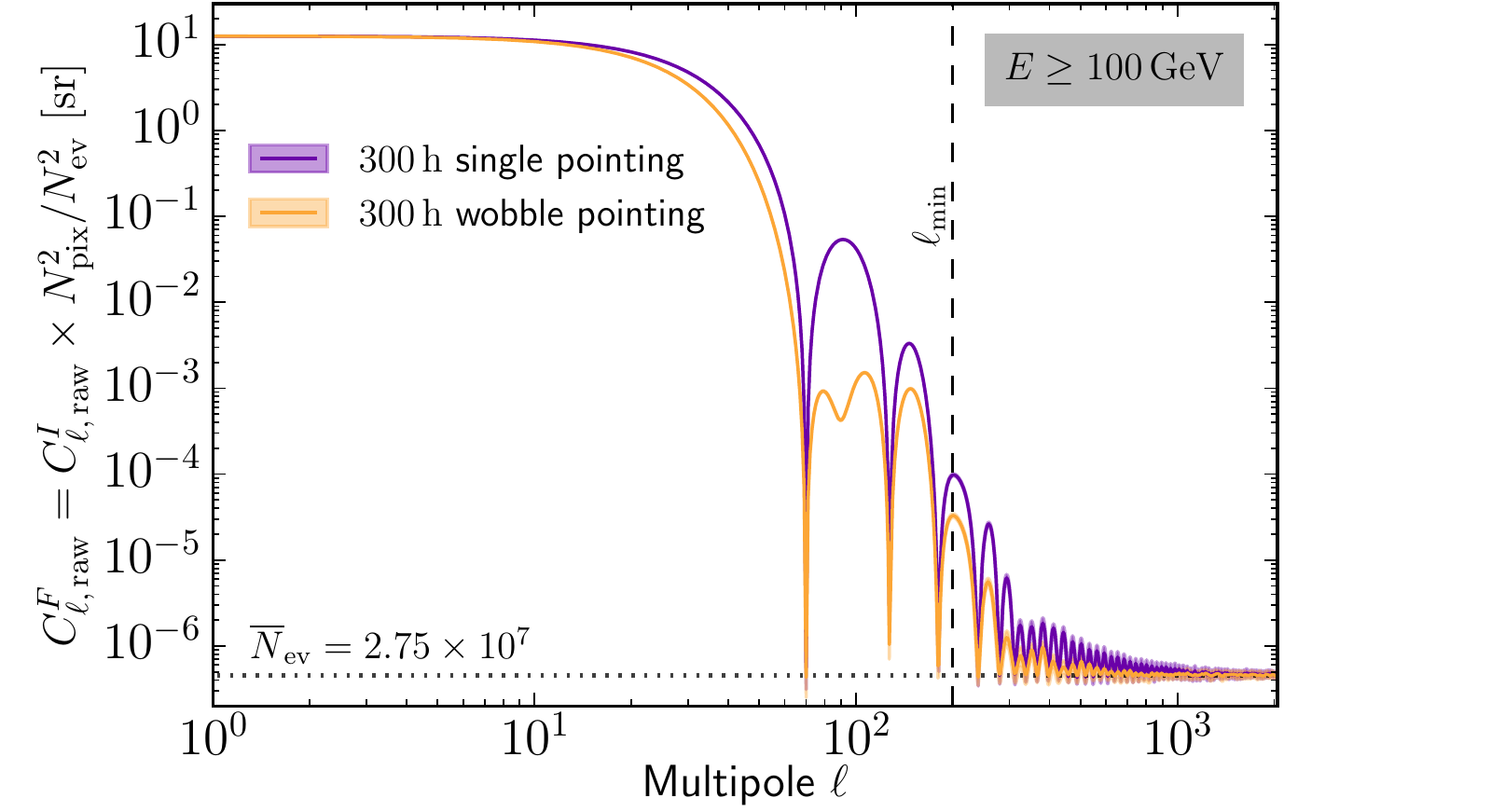}
\caption{
APS of all background events above $\unit[100]{GeV}$ in a single CTA pointed observation over  $\unit[300]{h}$ (for an instrumental response optimised for \unit[50]{h} of observation). We also show the APS for the $\unit[300]{h}$ observation time distributed over four observations, pointed $1.5\degs$ offset from the central pointing position (``wobble observation mode'').  The dotted line shows correspondingly $\fsky\times C_{\rm N}^F  = \frac{4\pi\,\unit{sr}}{\nevents{}}$. The vertical dashed line denotes the cut applicable for the quarter-sky survey APS (\cref{fig:APS_background_4in1}).
}
\label{fig:APS_background_singleFOV}
\end{center}
\end{figure}

\begin{figure}[t]
\begin{center}
\includegraphics[width=0.62\textwidth]{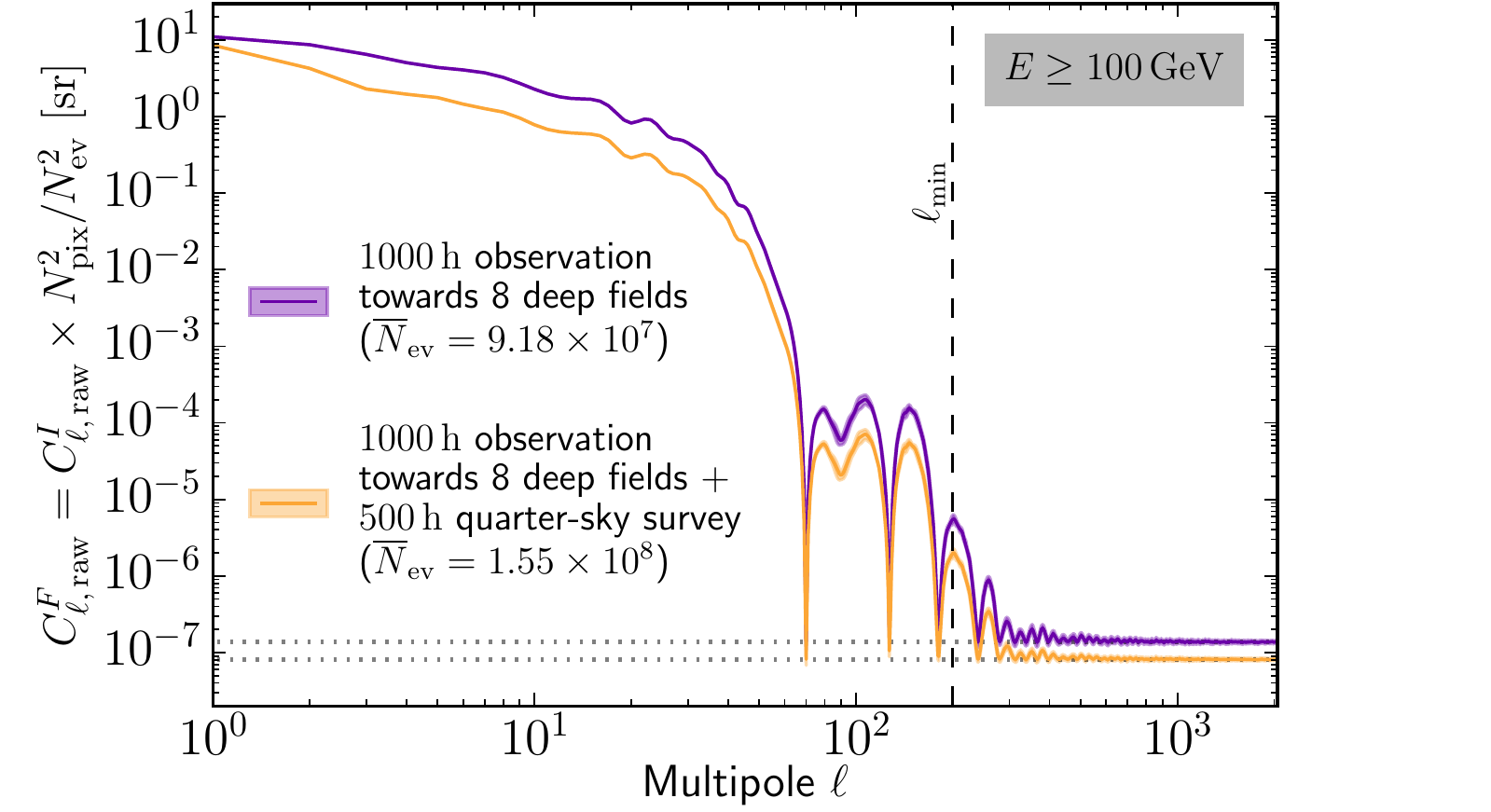}
\caption{
APS of all background events above $\unit[100]{GeV}$ towards eight different positions in the sky and a total observation time of $\unit[1000]{h}$ (for an instrumental response optimised for \unit[50]{h} of observation). We also show the APS for these observations combined with the events from a quarter-sky survey. 
}
\label{fig:APS_background_surveyDESdwarfs}
\end{center}
\end{figure}

The authors of \cite{2014JCAP...01..049R} also investigated whether a single or some few deep-exposure observations with IACTs can be used for an APS analysis. For their simplified Gaussian model of the CTA field of view (see \cref{fig:CTA_EffAreaOffAxis}), they found a slight increase in sensitivity when investing a fixed total observation time in a single deep field instead of distributing the time over observations of several fields. However, we argue that small field observation data may hardly compete with large-area survey data for a realistic, non-Gaussian CTA instrumental model: \Cref{fig:APS_background_singleFOV} shows the APS of the residual background above $\unit[100]{GeV}$ in a deep exposure ($\unit[300]{h}$) towards a single spot in the sky, where $C_{\ell,\,\rm raw}^F = \npix^2/\nevents^2 \times C_{\ell,\,\rm raw}^I$. The figure reveals dominant lobes in the spectrum  which are not present in the multipole transformation of a simple Gaussian function \cite{2014JCAP...01..049R}. The dashed vertical line again shows the corresponding $\ell_{\rm min}$ of our standard survey setup. Adopting a small set of several deep-exposure data sets does not much improve the situation, as we show in \cref{fig:APS_background_surveyDESdwarfs}. Here, $\unit[1000]{h}$ are distributed (in wobble observation mode) over eight pointings on the sky. The figure also shows that combining  deep-field observations with  survey data  does not attenuate the multipole lobes. For uncorrelated data sets, the joint APS of added data sets  is the sum of the individual spectra, and the oscillations are preserved. 

In principle, for known mask shapes, the spectral leakage can be eliminated from the spectrum by a rigorous calculation of $M_{\ell \ell' m m'}^{-1}$ from \cref{eq:Mllmm}, as done by, e.g.,  the \textsc{PolSpice} \cite{2001ApJ...548L.115S,2011ascl.soft09005C} or \textsc{Master} \cite{2002ApJ...567....2H} algorithms. However, even provided a perfect knowledge of the masking window in a given energy interval, the unmasked spectrum can only be reconstructed at the expense of noise amplification \cite{2012PhRvD..85h3007A}: For a mask being orders of magnitude larger than the physical signal in $C_{\ell}$ space, any information about the signal is likely to be buried in the shot noise and the systematic uncertainty about the mask. However, a rigorous investigation of unmasking small field-of-view \gr{} data is still to be done, e.g., whether the APS lobes can be suppressed by a suitable apodisation or tapering of the data. Such an advanced study would be particularly helpful to assess the usage of data from deep observations of dark spots (as, e.g., foreseen for the search for dark matter annihilation in dwarf spheroidal galaxies) for an APS analysis. For the remainder of this paper, we will focus on the APS analysis according to \cref{sec:aps_analysis}, applied to a wide-field CTA extragalactic survey with $\Delta_{\rm fov}=1\degs$.

\section{CTA sensitivity to anisotropies in the DGRB}
\label{sec:results}

\subsection{Sensitivity analysis description}

We  calculate the sensitivity to small-scale anisotropies in a CTA large-area survey data set at the 95\% confidence level (C.L.) detection threshold. This C.L. was also used by \cite{2014JCAP...01..049R}.  Therefore, we proceed as follows: We first generate mock skymaps, each containing $100\pi = 314$ point sources with equal flux level, ${\rm d}N/{\rm d}F = 100\pi/F_0\,\delta(F/F_0 -1)$, randomly distributed on the quarter-sky survey field. With $C_{\rm P}^I =  \frac{1}{4\pi\,\fsky\,\unit{sr}} \int F^2 \frac{\dd N}{\dd F}\,\dd F$ \cite{2009PhRvD..80b3520A}, such maps have an intrinsic  anisotropy of $C_{\rm P,\,input}^I = 100\,F_0^2\;\mathrm{sr}^{-1}= 10^{-2}\, I_\gamma^2\; \rm sr$  constant over all multipoles, where $I_\gamma = 100\,F_0\;\rm sr^{-1}$ is the intensity arising from these fluxes. Consequently, these sources generate a fixed fluctuation APS of $C_{\rm P,\,input}^F = C_{\rm P,\,input}^I/I_\gamma^2 \equiv \unit[10^{-2}]{sr}$. The flux of all the sources is modelled to follow the DGRB spectrum, \cref{eq:dgrb_ackermann2015}, $I_\gamma \propto I_{\rm DGRB}$. We use the {\tt ctobsim} ctool to simulate skymaps of the \gr{} and background events in the survey field and  we increase the normalisation of $I_\gamma$ until we detect an anisotropy signal, $\widehat{C}_{\rm P} > 0$, in the mock data at the $95\%$ C.L. with the likelihood test described in \cref{sec:aps_analysis}. According to \cref{eq:ts_distribution}, a $95\%$ C.L. detection  corresponds to a value $\rm{TS}\geq 2.71$ of our test statistic. We call the number of \gr{} events obtained from $I_\gamma$ at this sensitivity threshold $N_{\gamma,\,95\%}$. We repeat the calculation 25 times, varying the source positions on the sphere.\footnote{Note that by this procedure, we do not vary the intrinsic anisotropy of a constant intensity, but vary the intensity of a constant fluctuation anisotropy. By this, the total mean number of events, $\overline{N}_{\rm ev} = \overline{N}_{\rm bck} + \overline{N}_{\gamma}$, is not constant in our search for the sensitivity threshold. However, as $\overline{N}_{\gamma}< 10^{-4}\, \overline{N}_{\rm bck}$ and moreover $\Delta \overline{N}_{\gamma} \ll \sigma_{\nbck}$ in the scanned range of anisotropy levels (starting from $C_{\rm P,\,input}^F = \unit[10^{-2}]{sr}$), this difference in the approach is negligible.}

We perform our analysis in the four energy bins between 30 and 100 GeV, 100 and 300 GeV, 300 GeV and 1 TeV, and between 1 and 100 TeV; additionally we report the integrated sensitivity using all events above 30 GeV. As we collect all events in these energy bins, disregarding their individual energies, our results are virtually independent of the exact shape of the used DGRB spectrum. The only minor dependence of our analysis to the hypothesised DGRB spectral shape enters via the modelling of the average angular resolution in each energy bin, \cref{eq:avg_psf}. However, we consider this dependence negligible for the rescaling of the results between different DGRB spectra in the next subsection. 

\begin{figure}[t]
  \begin{center}
    \includegraphics[width=\textwidth]{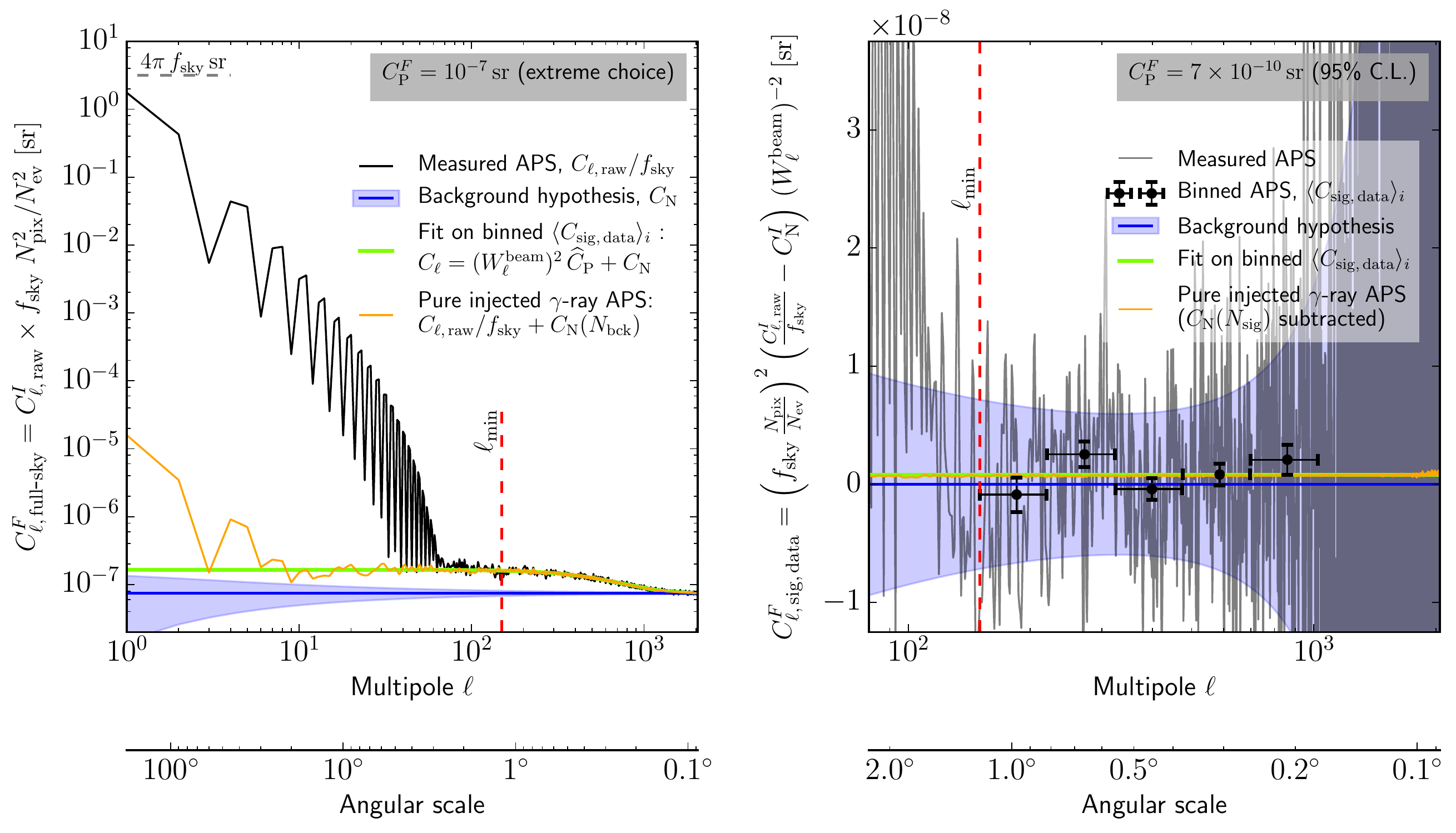}
  \captionof{figure}[Illustration of the APS likelihood fit at the 95\% C.L.
sensitivity threshold]{
Illustration of the APS likelihood fit in the energy bin $[100,\,300]\,\unit{GeV}$. {\it Left:} Global spectrum for an extreme fluctuation of the DGRB. It can be seen how the DGRB events (orange curve) add power on the top of the shot noise from the cosmic-ray background (blue), which however is obfuscated in the data by the sky mask at $\ell \lesssim 100$ (black curve). At $\ell \gtrsim 500$, the instrumental angular resolution suppresses the intrinsic signal power to the overall shot noise level. We evaluate the spectrum at $\ell_{\rm min}\leq \ell\leq 1024$. The green curve shows the likelihood fit to the data. It can be seen that the spectrum of simulated \gr{} events alone (orange curve) is well estimated and recovered by the fit. {\it Right:} Closeup of the interval $80\leq \ell \leq 2048$ for a signal APS at the 95\% C.L.
sensitivity threshold, with the average shot noise power already subtracted and the beam suppression undone.
}
  \label{fig:LikeIllustration_sensThreshold_paper}
  \end{center}
\end{figure}

\Cref{fig:LikeIllustration_sensThreshold_paper} illustrates the analysis and the equations from \cref{sec:aps_analysis}. In the left panel, we show an APS containing \gr{} and background events, however, for an extreme anisotropy level for illustration purpose only. To scrutinise our analysis procedure, we also show a ``pure'' spectrum from \gr{} events only with no cosmic-ray background (orange curve). This information is of course not available in a later analysis of real data. In the right panel, we show the APS after subtracting the shot noise and after unfolding (``signal APS''). The blue band indicates the expected  unbinned error, $\sigma_{\ell}(C_{\ell,\,\rm sig}=0)$  for the background-only hypothesis. This variance band is based on the  scaling $\sigma_{\ell,\, \rm raw} \approx 1.11\times\sigma_{\ell,\, \text{full-sky}}<1/\sqrt{\fsky}\;\sigma_{\ell,\, \text{full-sky}}$ empirically derived from a set of MC simulations.\footnote{See \cite{2015MNRAS.448.2854C} for a comprehensive study of the APS variance of non-Gaussian fields.} The black crosses in the right panel show the signal APS
 with the error, \cref{eq:APS_errorterm_fsky_binned}, after binning and maximising the likelihood. Finally, the green curve shows  the  likelihood fit to the binned ${C_{\ell,\,\rm sig}}$ and in both panels well agrees with the anisotropy of the  injected $N_\gamma$ \gr{} events (orange curves).

\subsection{Results}

In \cref{tab:cf_sensitivity}, we present the 95\% C.L. sensitivity  to small-scale anisotropies  in a CTA large-area survey,
\beq
C_{\rm P,\,95\%}^F= C_{\rm P,\, input}^F \times (N_{\gamma,\,95\%}/\nevents)^2\,.
\label{eq:CP95}
\eeq
 In the top panel, we show the anisotropies in the total data set (including all background and DGRB events) to which CTA is sensitive to, independent of the origin of these anisotropies. While we have carefully excluded expected artefacts from the analysis in \cref{sec:aps_analysis_background}, in principle, if any instrumental or physical origin generates anisotropies of these magnitude and on scales below $1.5\degs$, they are seen by our analysis. We will later discuss in \cref{sec:discussion} how this sensitivity can be interpreted with respect to small-scale cosmic-ray anisotropies. 

\begin{table}[t]
  \begin{flushleft}
  \resizebox{0.662\textwidth}{!}{%
  \begin{tabular}{|c|c c|} \hline
    \multicolumn{3}{| c |}{\cellcolor[gray]{0.8}  Sensitivity to anisotropies in all $\nevents$ events of the survey} \\ \hline
 Energy  & ``Measured'' total & Fluctuation sensitivity,\\
interval & events, $N_{\rm ev}$ &  $C_{\rm P,\,95\%}^F\;[\rm sr]$ \\\hline
&&\\[-0.3cm]
$[30,\,100]\,\mathrm{GeV}$               & $4.99\times 10^7$ & $(1.0\pm 0.7)\times 10^{-9}\;\,$  \\[0.2cm]
$[100,\,300]\,\mathrm{GeV}$              & $4.21\times 10^7$ & $(7.4\pm 4.3)\times 10^{-10}$  \\[0.2cm]
$[300,\,1000]\,\mathrm{GeV}$             & $1.26\times 10^7$ & $(2.1\pm 0.9)\times 10^{-9}\;\,$ \\[0.2cm]
$[1,\,100]\,\mathrm{TeV}$                & $8.51\times 10^6$ & $(2.8\pm 1.2)\times 10^{-9}\;\,$ \\[0.1cm]\hline
&&\\[-0.3cm]
$[30\,\mathrm{GeV},\,100\,\mathrm{TeV}]$ & $1.13\times 10^8$ & $(3.1\pm 2.1)\times 10^{-10}\,$ \\[0.1cm]\hline
  \end{tabular}
  }
  \end{flushleft}

 \resizebox{\textwidth}{!}{%
  \begin{tabular}{|c|c c|c c|} \hline
   \multicolumn{5}{| c |}{\cellcolor[gray]{0.8}  Sensitivity to anisotropies in a presumed DGRB intensity (top panel rescaled)} \\ \hline
 Energy  & Presumed events   & Sensitivity to        & Presumed events   & Sensitivity to         \\
interval & \nsig, \cref{eq:dgrb_ackermann2015} &  $C^F_{\rm P,\, DGRB}\;[\rm sr]$  & \nsig, \cref{eq:dgrb_brokenPL} &  $C^F_{\rm P,\, DGRB}\;[\rm sr]$  \\\hline
&&&&\\[-0.3cm]
$[30,\,100]\,\mathrm{GeV}$               & $10800$ & $(2.2\pm 1.5)\times 10^{-2}$ & $11500$  & $(2.0\pm 1.3)\times 10^{-2}$ \\[0.2cm]
$[100,\,300]\,\mathrm{GeV}$              & $14600$ & $(6.2\pm 3.6)\times 10^{-3}$ & $13100$  & $(7.7\pm 4.5)\times 10^{-3}$ \\[0.2cm]
$[300,\,1000]\,\mathrm{GeV}$             & $2660$  & $(4.8\pm 2.1)\times 10^{-2}$ & $4400$   & $(1.8\pm 0.8)\times 10^{-2}$ \\[0.2cm]
$[1,\,100]\,\mathrm{TeV}$                & $20$    & $> 4\pi\,f_{\rm sky}$        & $1300$   & $(1.2\pm 0.5)\times 10^{-1}$ \\[0.1cm]\hline
&&&&\\[-0.3cm]
$[30\,\mathrm{GeV},\,100\,\mathrm{TeV}]$ & $28100$ & $(5.0\pm 3.5)\times 10^{-3}$ & $30300$ & $(4.3\pm 3.0)\times 10^{-3}$ \\[0.1cm]\hline
  \end{tabular}
  }
 \caption[]{
{\it Top panel:} 95\% C.L.
sensitivity to any anisotropies in all $\nevents=\nbck+\nsig$ events of a $T_{\rm obs}=\unit[475]{h}$ CTA extragalactic survey data set. We report the median sensitivities $C_{\rm P,\,95\%}^F$ and, because of our analysis' limited statistic, a symmetric 68\% C.I.
{\it Bottom panel:} Sensitivities to small-scale anisotropies in the DGRB for an exponentially suppressed DGRB, \cref{eq:dgrb_ackermann2015} (left columns) or a broken power-law extrapolation \cref{eq:dgrb_brokenPL} (right columns).
  }
\label{tab:cf_sensitivity}
\end{table}

In a second step, we  assume that all anisotropies are generated by the \grs{} of the DGRB. Taking the results from the top panel of \cref{tab:cf_sensitivity} and some additional knowledge about the average DGRB intensity,  \cref{eq:APS_subsourceclass} can be used to draw a statement about the relative fluctuation level in the DGRB detectable by our analysis. This is presented in the lower panel of \cref{tab:cf_sensitivity}. Here, we show this sensitivity assuming an exponentially suppressed DGRB intensity, \cref{eq:dgrb_ackermann2015}, on the left.  While \cite{2015ApJ...799...86A} have excluded  a pure power-law scaling of the DGRB in the VHE regime at the $5.7\sigma$ level, they cannot distinguish between an exponential cut-off of the DGRB (suggested by the absorption of distant \grs{} on the EBL) and a continuation of the spectrum with a steeper power-law slope (which would hint to a different source class  at the highest energies). Therefore, we additionally investigate our results for a broken power-law DGRB,
\beq
I_{\text{DGRB}} = \begin{cases}
I_{\rm b}\,(E/E_{\rm b})^{-\gamma_1}\quad\text{if}\; E<E_{\rm b}\\
I_{\rm b}\,(E/E_{\rm b})^{-\gamma_2}\quad\text{otherwise,}
\end{cases}
\label{eq:dgrb_brokenPL}
\eeq
with $I_{\rm b}=\unit[5.74\times 10^{-14}]{cm^{-2}\,s^{-1}\,sr^{-1}\,MeV^{-1}}$, $\gamma_1=2.30$, $\gamma_2=3.07$, and $E_{\rm b}=\unit[55.3]{GeV}$ obtained from a fit to the \fermi{} data from \cite[model B]{2015ApJ...799...86A}. The APS sensitivity for such a broken power-law DGRB intensity is shown in the right columns of \cref{tab:cf_sensitivity}. 
It can be seen that a power-law behaviour of the DGRB is needed in the TeV regime for generating detectable anisotropies, $C^F_{\rm P}\leq 4\pi\,f_{\rm sky}$ above $\unit[1]{TeV}$. The CTA sensitivity assuming a broken power-law extrapolation of the DGRB is finally also displayed in \cref{fig:CP-Energy}. We stress that an interpretation of a detected anisotropy power in terms of the DGRB relies on complementary measurements of its diffuse intensity and the corresponding expected number of CTA events.

Our results of the CTA sensitivity  can be compared to   the previous study from \cite{2014JCAP...01..049R} using $C^F_{\rm P,\, DGRB} = (1 - f_{\rm DM} )^2 \times C^F_{\rm P,\, astro} + f_{\rm DM}^2 \times C^F_{\rm P,\, DM}$. The authors of \cite{2014JCAP...01..049R} fixed $C^F_{\rm P,\, astro}\equiv \unit[10^{-5}]{sr}$, $C^F_{\rm P,\, DM}\equiv \unit[10^{-3}]{sr}$ and expressed their sensitivity in terms of the detectable fraction $f_{\rm DM}$ of dark matter induced anisotropies. With a detectable $C^F_{\rm P,\, DGRB}\gtrsim \unit[4\times 10^{-3}]{sr} $, we find our result comparable to their most conservative assumptions about the CTA performance ($f_{\rm DM} \gtrsim 1$).\footnote{Independent of the presumed DGRB spectrum, the CTA configuration $E_{\rm th}=\unit[300]{GeV}$, an observation time of $10\times \unit[100]{h}$, and $\unit[10]{Hz}$ background rate from \cite{2014JCAP...01..049R} yields the sensitivity to $C^F_{\rm P}= 1\times\unit[10^{-9}]{sr}$ in all events, most similar to our analysis ($C_{\rm P,\,95\%}^F=\unit[2.1\times 10^{-9}]{sr}$ between \unit[300]{GeV} and \unit[1]{TeV}, see upper panel of \cref{tab:cf_sensitivity}).}

\section{Discussion}
\label{sec:discussion}

\paragraph{DGRB anisotropies measured by the \fermi{}:}

\begin{figure}[t]
  \begin{center}
    \includegraphics[width=0.8\textwidth]{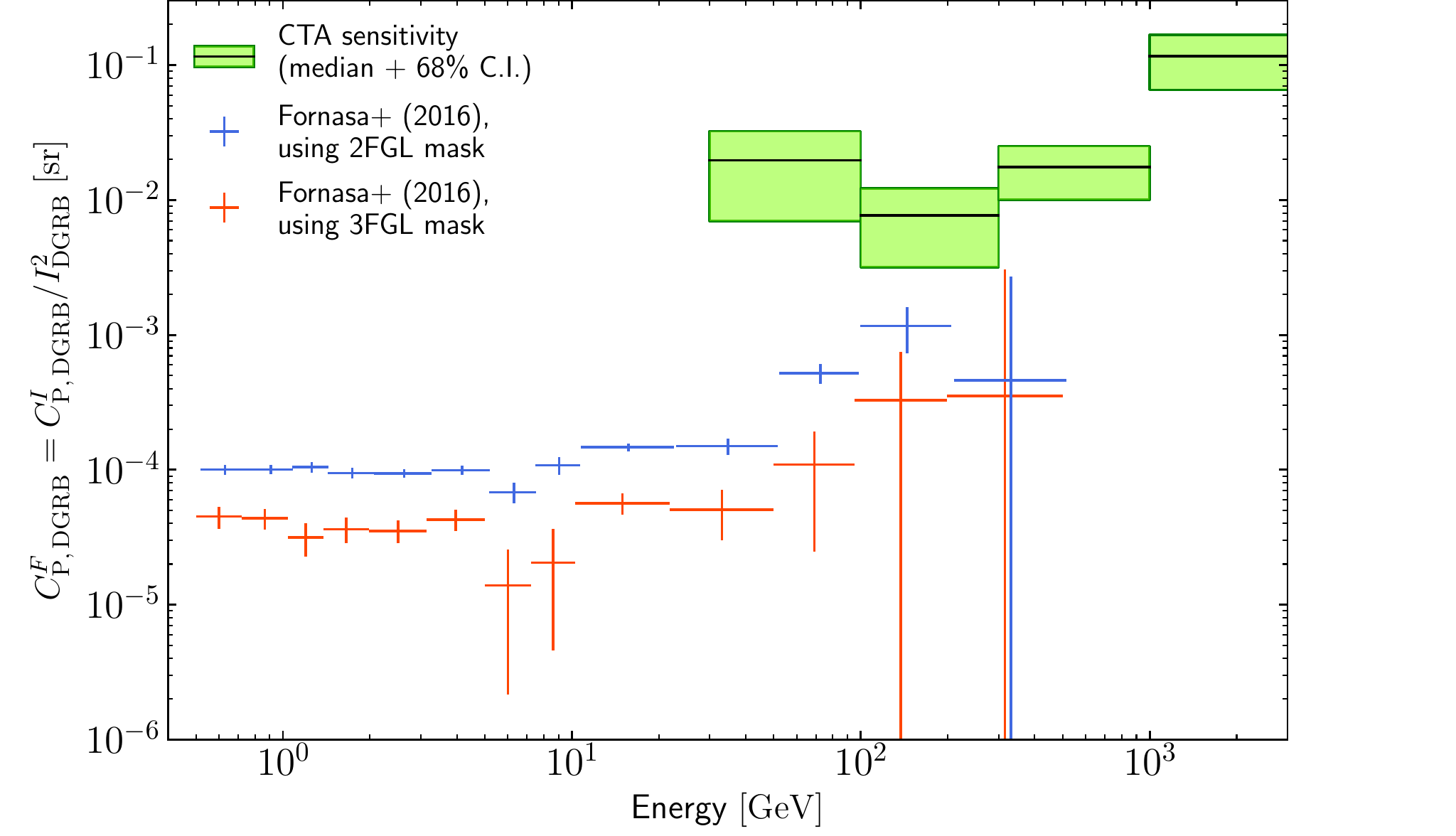}
  \captionof{figure}[Sensitivity of the CTA extragalactic survey to DGRB anisotropies compared to the most recent measurement by \textit{Fermi}-LAT]{
Sensitivity of the CTA extragalactic survey to anisotropies in the DGRB compared to the measurement by the \fermi{} \citep{2016PhRvD..94l3005F}. The figure shows the relative anisotropy levels, $C^F_{\rm P,\,DGRB}$, with respect to the broken power-law  DGRB intensity, \cref{eq:dgrb_brokenPL}. The highest energy bin for the CTA sensitivity shows the  sensitivity integrated   up to $\unit[100]{TeV}$. 
}
  \label{fig:CP-Energy}
  \end{center}
\end{figure} 

\begin{figure}[t]
\centering
\includegraphics[width=\textwidth]{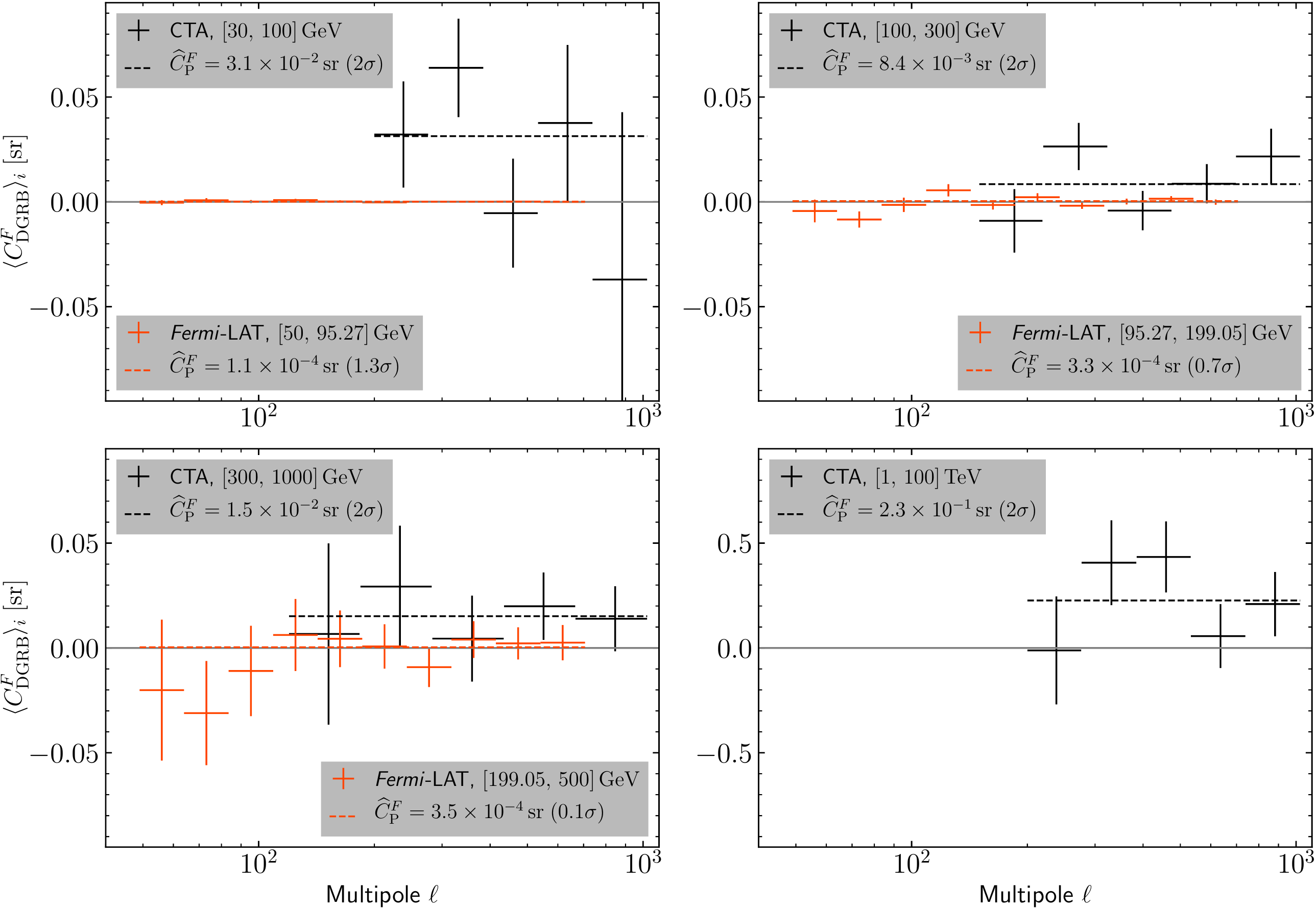}
\caption{Comparison of the binned APS in the three overlapping bins between the \fermi{} analysis \cite{2016PhRvD..94l3005F} (masking 3FGL sources) and our CTA sensitivity calculation at the 95\% C.L. As in \cref{fig:CP-Energy}, we show the fluctuation APS with respect to the DGRB intensity of \cref{eq:dgrb_brokenPL}. Note that the CTA points and the fit in the upper right panel are the same as shown in \cref{fig:LikeIllustration_sensThreshold_paper} (right). Also note the larger vertical axis scale in the lower right panel.
}
\label{fig:APS_comp_Fermi_rel}
\end{figure}

Small-scale anisotropies in the DGRB have already been detected in the data from the \fermi{}. Taking 81 months of reprocessed \texttt{Pass7} data between \unit[500]{MeV} and \unit[500]{GeV}, the authors of \cite{2016PhRvD..94l3005F} have recently updated the earlier result from \cite{2012PhRvD..85h3007A}, based on 22 months of \fermi{} data. In addition to covering a larger energy range and better precision compared to the first measurement, \cite{2016PhRvD..94l3005F} conclude on two power-law populations of point sources dominating the DGRB anisotropies below and above \unit[2]{GeV}, respectively. To compare with our results, in \cref{fig:CP-Energy}, we show the \fermi{} measurement of the auto-correlation APS in terms of the fluctuation APS, $C_{\rm P,\,DGRB}^F$,  relative to the average DGRB intensity of \cref{eq:dgrb_brokenPL}.\footnote{Note that in \cref{fig:CP-Energy}, the \textit{relative location} of the measured  data points and the CTA sensitivity in terms of the fluctuation APS is independent of the assumed $I_{\rm DGRB}$.} The authors of \cite{2016PhRvD..94l3005F} studied the APS for two slightly different data sets, excluding all 3FGL sources (orange crosses in \cref{fig:CP-Energy}) and only excluding the 2FGL sources (blue crosses). As expected, the DGRB power is reduced when masking the power originating from additionally resolved sources in the 3FGL. From the comparison between the \fermi{} measurement and our results, it becomes evident that the existing detection of DGRB anisotropies from \cite{2016PhRvD..94l3005F} already weakly excludes the small-scale DGRB anisotropies accessible with CTA. However, the \fermi{} measurement is based on a slightly lower multipole range ($49\leq \ell \leq 706$). Also, the statistical uncertainty of the \fermi{} measurement in its highest energy bins is large; and in the regime of saturation of the \fermi{} detector, systematic biases may additionally increase the uncertainty. This is illustrated in \cref{fig:APS_comp_Fermi_rel} by a more close comparison between the three overlapping energy bins of the \fermi{} measurement \cite{2016PhRvD..94l3005F} and our CTA sensitivity. Notably, a potential oversubtraction of the mask in the lowest $\ell$-bins in the \fermi{} measurement may artificially reduce the detected signal. We also remark that the energy APS from \cite{2016PhRvD..94l3005F} shows low-significant features which are not yet fully understood, e.g., a peak of the angular power in the single energy bin between 50 and \unit[95.27]{GeV} when using \texttt{Pass8} data (which is not present for the default analysis relying on \texttt{Pass7} data). Therefore, we conclude that the reach of small-scale \gr{} anisotropies with CTA is not ultimately excluded by \fermi{} results, in particular at the highest energies above \unit[300]{GeV}, and a complementary analysis with CTA may help to reduce instrument-related systematic uncertainties.

\paragraph{Expected VHE anisotropy levels of various source classes:}

Several estimates have been drawn to assess the expected anisotropy level from unresolved members of various source classes contributing to the DGRB \cite{2011MNRAS.415.1074S,2014JCAP...11..021D,2016JCAP...09..047H,2007PhRvD..75f3519A,2013MNRAS.429.1529F,2009MNRAS.400.2122A}. While this has been already extensively discussed by \cite{2012PhRvD..85h3007A,2015PhR...598....1F}, we present a short overview of these estimates together with our calculation of the CTA sensitivity in  \cref{tab:cp_astro}.  For the APS signal from Galactic DM substructure, we rely on our  previous work \cite{2016JCAP...09..047H}, where we have comprehensively bracketed the expected intensity and APS characteristics from Galactic DM. While we have presented in figure~7 of \cite{2016JCAP...09..047H} the dimensional intensity APS for a particular particle physics model, we report in \cref{tab:cp_astro} $C^F_{\rm \ell=100} = C^I_{\rm \ell=100}/\langle I_{\rm DM}\rangle^2$  with $\langle I_{\rm DM}\rangle$  the average emission from all Galactic DM (smooth halo and subhalos) in our definition of the CTA extragalactic survey field.

Using these estimates, the second to last column in  \cref{tab:cp_astro}  indicates the fraction of intensity of the respective source class to the DGRB to create anisotropies detectable with CTA. This is confronted with a coarse estimate of the predicted DGRB fractions (last column of \cref{tab:cp_astro}). It is worth noting that  \cite{2016PhRvL.116o1105A} claim that in the VHE range, $E>\unit[50]{GeV}$, the DGRB is dominated by the emission from unresolved blazars with $f_{\rm DGRB}=86^{+16}_{-14}\%$, although their anisotropy imprint may be too low to be detectable with CTA. A signal from Galactic DM substructure would only bear the chance of a detection for the most optimistic clustering model from \cite{2016JCAP...09..047H} and the full DGRB (at the peak of the DM-induced \gr{} spectrum) originating from DM annihilation. This would apply to annihilation cross sections of $\sigmav \gtrsim \unit[10^{-23}]{cm^3\,s^{-1}}$ for DM particle masses $\mchi{}\sim \unit[1]{TeV}$ and annihilation into bottom quarks. Although this suggests that the indirect search for DM signals with CTA via \gr{} anisotropies is not very competitive, we stress that this conclusion only applies for the standard paradigm of weakly interacting massive dark matter particles in a $\Lambda$CDM Universe and only undergoing DM self-interactions. Moreover, a dedicated study tailored to spatially extended Galactic and extragalactic DM structures is excluded from  this work. Overall, we finally conclude that an APS analysis with CTA is most promising to constrain the contribution of misaligned AGN (as already pointed out by \cite{2014JCAP...11..021D}) and possibly also unresolved blazars to the DGRB in the VHE regime.

\begin{table}[!t]
\centering
\resizebox{\textwidth}{!}{%
 \begin{tabular}{|c | c c |c c c |} \hline
\multirow{2}{*}{Source class}   & \multicolumn{2}{c|}{Predicted}                              & Detectable $f_{\rm DGRB}$ & \multicolumn{2}{c|}{\multirow{2}{*}{Predicted $f_{\rm DGRB}$}} \\
                                & \multicolumn{2}{c|}{$C^F_{\rm \ell=100}\;[\rm sr]$}                & $(>30\,\rm GeV)$  & &\\\hline
Millisecond pulsars             & $3\times10^{-2}$       & \cite{2011MNRAS.415.1074S}            & 38\%              & $< 1\%$& \cite{2014ApJ...796...14C} \\ [0.1cm] 
Misaligned AGN                  &  \multicolumn{2}{l|}{$\;\,\gtrsim \;10^{-3}$ ($>\unit[100]{GeV}$) \cite{2014JCAP...11..021D}  }             & $\sim 1$ & $\lesssim 25\%$& \cite{2014ApJ...780..161D}\\[0.1cm] 
Galactic DM substructure (HIGH) & $3\times10^{-4}$       & \cite[this work]{2016JCAP...09..047H} & $\gtrsim 1$& \multicolumn{1}{c}{depending on \sigmav}&\\[0.1cm] 
Unresolved blazars              & $2\times10^{-4}$       & \cite{2007PhRvD..75f3519A} &$\gtrsim 1$      & $\gtrsim 80\%$ $(>50\,\rm GeV)$& \cite{2016PhRvL.116o1105A} \\[0.1cm] 
Galactic DM substructure (LOW)  & $3\times10^{-5}$       & \cite[this work]{2016JCAP...09..047H} &$>1$ & \multicolumn{1}{c}{depending on \sigmav}& \\[0.1cm]  
Extragalactic DM structure   & $\gtrsim \;\;10^{-5}$  & \cite{2013MNRAS.429.1529F}            &$>1$ & \multicolumn{1}{c}{depending on \sigmav}& \\[0.1cm]  
Star-forming galaxies           & $2\times10^{-7}$       & \cite{2009MNRAS.400.2122A}            &$>1$ & $\lesssim 50\%$& \cite{2014JCAP...09..043T}\\[0.1cm] 
\hline
 \end{tabular}
} 
\caption[APS of unresolved contributors from different source classes]{Predicted contributions to the DGRB of various unresolved source classes. If not explicitly stated, all quoted values for $C^F_\ell$ and $f_{\rm DGRB}$ are assumed constant in the $\gtrsim \unit[100]{MeV}$ range. For DM annihilation, the predicted intensity depends on the velocity-averaged annihilation cross-section, \sigmav{}.
}
\label{tab:cp_astro}
\end{table}

\paragraph{Are small-scale cosmic-ray anisotropies excluded?}

Throughout this paper, we have assumed a perfectly isotropised CTA residual background of electrons and misclassified hadrons. Due to the large relative amount of these background events, however, even tiny amplitudes of $C^F_{\rm bck} \gtrsim \unit[10^{-9}]{sr}$ in the background could dominate over any \gr{} anisotropies. In fact, cosmic-ray anisotropies at scales $\gtrsim 5\degs$ are since long known \cite{2005ApJ...626L..29A,2009ApJ...698.2121A}, with  \textit{hadronic} cosmic-ray  anisotropies at the level of $C^F_{\ell} \lesssim \unit[10^{-9}]{sr}$ up to $\ell \lesssim 40$ \cite{2015arXiv151004134T,2017arXiv170803005T}. However, even when extrapolating this level beyond $\ell\gtrsim 100$, given that less than 1\% of all hadrons survive the CTA $\gamma$-hadron separation cuts, remaining hadronic cosmic-ray anisotropies in the CTA background are smaller than $C^F_{\ell} \lesssim \unit[10^{13}]{sr}$ and thus negligible.

The \fermi{} collaboration has recently published a search for anisotropies in the  \textit{electron intensity} based on almost seven years of data  \cite{2017PhRvL.118i1103A}. No   electron anisotropy has been found, with an energy-dependent upper limit on the dipole anisotropy of $C_{\ell=1,\,95\%\rm C.L.}^F\lesssim \unit[10^{-2}]{sr}$. As their analysis suggests a sensitivity to $C^F_{\ell} \lesssim \unit[10^{-6}]{sr}$ at $\ell\geq 100$ and $E\geq \unit[100]{GeV}$,\footnote{See their figure~11 in the supplementary material to \cite{2017PhRvL.118i1103A}.}   it remains to comprehensively study whether  CTA's sensitivity to small-scale anisotropies may supersede \fermi{}'s ability to measure electron anisotropies at $\ell\gtrsim 100$.

However, although low-amplitude small-scale electron anisotropies may be within CTA's sensitivity range, we judge their existence to be highly unlikely. Only primary cosmic electrons, accelerated in point-like sources and with sufficiently large Larmor-radii, are able to generate small-scale anisotropies. Because of their fast synchrotron cooling, such primary electrons must  originate  from nearby ($d\lesssim \unit[1]{kpc}$) sources, and only a few supernova remnants constitute plausible candidates \cite{2015PhLB..749..267L}. Also, the scenario of DM annihilation in nearby subhalos generating electron anisotropies is considered to be already ruled out \cite{2012APh....35..537B,2015JCAP...02..043P}.

\section{Conclusions}
\label{sec:conclusions}

In this work, we have studied whether CTA is able to probe small-scale anisotropies in the DGRB, based on the latest knowledge about the future instrument. Using standard background rejection cuts and a binned likelihood analysis, we have found a sensitivity to  \gr{} anisotropies with relative amplitudes of $C^F_{\rm P,\,DGRB} \gtrsim 4\times \unit[10^{-3}]{sr}$  above 30 GeV and scales smaller than $1.5\degs$. Beyond $\unit[1]{TeV}$, such anisotropies may  only be detectable if the DGRB is not exponentially suppressed at sub-TeV energies.  While small-scale anisotropies are most likely solely generated by \grs{}, our results can also be interpreted in terms of any anisotropies present in the data, to which our results show a sensitivity to $C^F_{\rm P} \gtrsim 3\times \unit[10^{-10}]{sr}$.  

We have confronted this sensitivity to previous detections of DGRB anisotropies in the \fermi{} data. CTA may be in reach to detect the anisotropy levels found at low significance by the \fermi{}, in particular at \gr{} energies above \unit[300]{GeV}. A complementary analysis of CTA data may also help to rule out systematic biases affecting the current and future \cite{2017ifs..confE.158N} \fermi{} measurements. Discussing various classes of unresolved astrophysical  \gr{} emitters, a CTA analysis of VHE \gr{} anisotropies will be primarily able to probe current models for the population of unresolved blazars and  misaligned AGN.

Investigating various anticipated CTA data sets, we have found the CTA extragalactic survey to be most suitable for this kind of analysis; and the analysis presented in this paper proposes an additional way of studying the survey data. On the other hand, we have stated that the analysis of single-FOV data for a non-Gaussian  acceptance of the instrument is heavily impeded by artefacts in the multipole transformation. Discussing various sources of pollution to an ideal survey data set, we have obtained that a homogeneous exposure is most crucial for the success of the search for small-scale anisotropies. We therefore recommend a survey tiling as small as $1\degs$ or even a slew survey, where data is taken continuously. Whether the sensitivity of CTA to \gr{} anisotropies can be further improved by such a slew survey (at the potential expense of a worse angular resolution), using divergent telescope pointing, dedicated analysis cuts, or a sophisticated modelling of the background APS, is finally left to future work.

\acknowledgments

This work was conducted in the context of the CTA Extragalactic and Fundamental Physics  Working Groups and has gone through internal review by the CTA Consortium; we thank the internal reviewers Josep Mart\'i and Germ\'an G\'omez-Vargas for their valuable remarks. Also, we  thank Elisa Pueschel for her comments that helped to substantially improve the quality of the manuscript. This research has made use of the CTA instrument response functions provided by the CTA Consortium and Observatory, see \url{http://www.cta-observatory.org/science/cta-performance} (version \texttt{prod3b-v1}) for more details.  This work was supported by the Research Training Group 1504, ``Mass, Spectrum, Symmetry'', of the German Research Foundation (DFG) and the Max Planck Society (MPG).  Some of the results in this article have been derived using the \healpix{} package \cite{2005ApJ...622..759G}.

\bibliography{APSpaper}

\providecommand{\href}[2]{#2}\begingroup\raggedright\begin{thebibliography}{10}

\bibitem{2009ApJ...697.1071A}
W.~B. {Atwood}, A.~A. {Abdo}, M.~{Ackermann}, W.~{Althouse}, B.~{Anderson},
  M.~{Axelsson} et~al., \emph{{The Large Area Telescope on the Fermi Gamma-Ray
  Space Telescope Mission}},
  \href{http://dx.doi.org/10.1088/0004-637X/697/2/1071}{\emph{\apj} {\bf 697}
  (June, 2009) 1071--1102}, [\href{http://arxiv.org/abs/0902.1089}{{\tt
  0902.1089}}].

\bibitem{2015ApJS..218...23A}
F.~{Acero}, M.~{Ackermann}, M.~{Ajello}, A.~{Albert}, W.~B. {Atwood},
  M.~{Axelsson} et~al., \emph{{Fermi Large Area Telescope Third Source
  Catalog}}, \href{http://dx.doi.org/10.1088/0067-0049/218/2/23}{\emph{\apjs}
  {\bf 218} (June, 2015) 23}, [\href{http://arxiv.org/abs/1501.02003}{{\tt
  1501.02003}}].

\bibitem{2008ICRC....3.1341W}
S.~P. {Wakely} and D.~{Horan}, \emph{{TeVCat: An online catalog for Very High
  Energy Gamma-Ray Astronomy}}, {\emph{International Cosmic Ray Conference}
  {\bf 3} (2008) 1341--1344}.

\bibitem{1972ApJ...177..341K}
W.~L. {Kraushaar}, G.~W. {Clark}, G.~P. {Garmire}, R.~{Borken}, P.~{Higbie},
  V.~{Leong} et~al., \emph{{High-Energy Cosmic Gamma-Ray Observations from the
  OSO-3 Satellite}}, \href{http://dx.doi.org/10.1086/151713}{\emph{\apj} {\bf
  177} (Nov., 1972) 341}.

\bibitem{1975ApJ...198..163F}
C.~E. {Fichtel}, R.~C. {Hartman}, D.~A. {Kniffen}, D.~J. {Thompson}, G.~F.
  {Bignami}, H.~{{\"O}gelman} et~al., \emph{{High-energy gamma-ray results from
  the second small astronomy satellite}},
  \href{http://dx.doi.org/10.1086/153590}{\emph{\apj} {\bf 198} (May, 1975)
  163--182}.

\bibitem{2012ApJ...750....3A}
M.~{Ackermann}, M.~{Ajello}, W.~B. {Atwood}, L.~{Baldini}, J.~{Ballet},
  G.~{Barbiellini} et~al., \emph{{Fermi-LAT Observations of the Diffuse
  {$\gamma$}-Ray Emission: Implications for Cosmic Rays and the Interstellar
  Medium}}, \href{http://dx.doi.org/10.1088/0004-637X/750/1/3}{\emph{\apj} {\bf
  750} (May, 2012) 3}, [\href{http://arxiv.org/abs/1202.4039}{{\tt
  1202.4039}}].

\bibitem{2010ApJ...724.1044S}
M.~{Su}, T.~R. {Slatyer} and D.~P. {Finkbeiner}, \emph{{Giant Gamma-ray Bubbles
  from Fermi-LAT: Active Galactic Nucleus Activity or Bipolar Galactic Wind?}},
  \href{http://dx.doi.org/10.1088/0004-637X/724/2/1044}{\emph{\apj} {\bf 724}
  (Dec., 2010) 1044--1082}, [\href{http://arxiv.org/abs/1005.5480}{{\tt
  1005.5480}}].

\bibitem{2009arXiv0912.3478C}
J.-M. {Casandjian}, I.~{Grenier} and {for the Fermi Large Area Telescope
  Collaboration}, \emph{{High Energy Gamma-Ray Emission from the Loop I
  region}}, {\emph{ArXiv e-prints} (Dec., 2009) \!\!},
  [\href{http://arxiv.org/abs/0912.3478}{{\tt 0912.3478}}].

\bibitem{2010PhRvL.104j1101A}
A.~A. {Abdo}, M.~{Ackermann}, M.~{Ajello}, W.~B. {Atwood}, L.~{Baldini},
  J.~{Ballet} et~al., \emph{{Spectrum of the Isotropic Diffuse Gamma-Ray
  Emission Derived from First-Year Fermi Large Area Telescope Data}},
  \href{http://dx.doi.org/10.1103/PhysRevLett.104.101101}{\emph{\prl} {\bf 104}
  (Mar., 2010) 101101}, [\href{http://arxiv.org/abs/1002.3603}{{\tt
  1002.3603}}].

\bibitem{2015ApJ...799...86A}
M.~{Ackermann}, M.~{Ajello}, A.~{Albert}, W.~B. {Atwood}, L.~{Baldini},
  J.~{Ballet} et~al., \emph{{The Spectrum of Isotropic Diffuse Gamma-Ray
  Emission between 100 MeV and 820 GeV}},
  \href{http://dx.doi.org/10.1088/0004-637X/799/1/86}{\emph{\apj} {\bf 799}
  (Jan., 2015) 86}, [\href{http://arxiv.org/abs/1410.3696}{{\tt 1410.3696}}].

\bibitem{2007AIPC..921..122D}
C.~D. {Dermer}, \emph{{The Extragalactic {$\gamma$}-ray Background}},  in
  \emph{The First GLAST Symposium} (S.~{Ritz}, P.~{Michelson} and C.~A.
  {Meegan}, eds.), vol.~921 of \emph{American Institute of Physics Conference
  Series}, pp.~122--126, July, 2007.
\newblock \href{http://arxiv.org/abs/0704.2888}{{\tt 0704.2888}}.
\newblock \href{http://dx.doi.org/10.1063/1.2757282}{DOI}.

\bibitem{2015PhR...598....1F}
M.~{Fornasa} and M.~A. {S{\'a}nchez-Conde}, \emph{{The nature of the Diffuse
  Gamma-Ray Background}},
  \href{http://dx.doi.org/10.1016/j.physrep.2015.09.002}{\emph{\physrep} {\bf
  598} (Oct., 2015) 1--58}, [\href{http://arxiv.org/abs/1502.02866}{{\tt
  1502.02866}}].

\bibitem{2014JCAP...09..043T}
I.~{Tamborra}, S.~{Ando} and K.~{Murase}, \emph{{Star-forming galaxies as the
  origin of diffuse high-energy backgrounds: gamma-ray and neutrino
  connections, and implications for starburst history}},
  \href{http://dx.doi.org/10.1088/1475-7516/2014/09/043}{\emph{\jcap} {\bf 09}
  (Sept., 2014) 043}, [\href{http://arxiv.org/abs/1404.1189}{{\tt 1404.1189}}].

\bibitem{2014ApJ...780..161D}
M.~{Di Mauro}, F.~{Calore}, F.~{Donato}, M.~{Ajello} and L.~{Latronico},
  \emph{{Diffuse {$\gamma$}-Ray Emission from Misaligned Active Galactic
  Nuclei}}, \href{http://dx.doi.org/10.1088/0004-637X/780/2/161}{\emph{\apj}
  {\bf 780} (Jan., 2014) 161}, [\href{http://arxiv.org/abs/1304.0908}{{\tt
  1304.0908}}].

\bibitem{2015PhRvD..91l3001D}
M.~{Di Mauro} and F.~{Donato}, \emph{{Composition of the Fermi-LAT isotropic
  gamma-ray background intensity: Emission from extragalactic point sources and
  dark matter annihilations}},
  \href{http://dx.doi.org/10.1103/PhysRevD.91.123001}{\emph{\prd} {\bf 91}
  (June, 2015) 123001}, [\href{http://arxiv.org/abs/1501.05316}{{\tt
  1501.05316}}].

\bibitem{2015ApJ...800L..27A}
M.~{Ajello}, D.~{Gasparrini}, M.~{S{\'a}nchez-Conde}, G.~{Zaharijas},
  M.~{Gustafsson}, J.~{Cohen-Tanugi} et~al., \emph{{The Origin of the
  Extragalactic Gamma-Ray Background and Implications for Dark Matter
  Annihilation}},
  \href{http://dx.doi.org/10.1088/2041-8205/800/2/L27}{\emph{\apjl} {\bf 800}
  (Feb., 2015) L27}, [\href{http://arxiv.org/abs/1501.05301}{{\tt
  1501.05301}}].

\bibitem{2016PhRvL.116o1105A}
M.~{Ackermann}, M.~{Ajello}, A.~{Albert}, W.~B. {Atwood}, L.~{Baldini},
  J.~{Ballet} et~al., \emph{{Resolving the Extragalactic {$\gamma$}-ray
  Background above 50 GeV with the Fermi Large Area Telescope}},
  \href{http://dx.doi.org/10.1103/PhysRevLett.116.151105}{\emph{\prl} {\bf 116}
  (Apr., 2016) 151105}, [\href{http://arxiv.org/abs/1511.00693}{{\tt
  1511.00693}}].

\bibitem{2018ApJ...856..106D}
M.~{Di Mauro}, S.~{Manconi}, H.-S. {Zechlin}, M.~{Ajello}, E.~{Charles} and
  F.~{Donato}, \emph{{Deriving the Contribution of Blazars to the Fermi-LAT
  Extragalactic {$\gamma$}-ray Background at $E>10$ GeV with Efficiency
  Corrections and Photon Statistics}},
  \href{http://dx.doi.org/10.3847/1538-4357/aab3e5}{\emph{\apj} {\bf 856}
  (Apr., 2018) 106}, [\href{http://arxiv.org/abs/1711.03111}{{\tt
  1711.03111}}].

\bibitem{2011MNRAS.415.1074S}
J.~M. {Siegal-Gaskins}, R.~{Reesman}, V.~{Pavlidou}, S.~{Profumo} and T.~P.
  {Walker}, \emph{{Anisotropies in the gamma-ray sky from millisecond
  pulsars}},
  \href{http://dx.doi.org/10.1111/j.1365-2966.2011.18672.x}{\emph{\mnras} {\bf
  415} (Aug., 2011) 1074--1082}, [\href{http://arxiv.org/abs/1011.5501}{{\tt
  1011.5501}}].

\bibitem{2014ApJ...796...14C}
F.~{Calore}, M.~{Di Mauro} and F.~{Donato}, \emph{{Diffuse {$\gamma$}-Ray
  Emission from Galactic Pulsars}},
  \href{http://dx.doi.org/10.1088/0004-637X/796/1/14}{\emph{\apj} {\bf 796}
  (Nov., 2014) 14}, [\href{http://arxiv.org/abs/1406.2706}{{\tt 1406.2706}}].

\bibitem{1978ApJ...223.1015G}
J.~E. {Gunn}, B.~W. {Lee}, I.~{Lerche}, D.~N. {Schramm} and G.~{Steigman},
  \emph{{Some astrophysical consequences of the existence of a heavy stable
  neutral lepton}}, \href{http://dx.doi.org/10.1086/156335}{\emph{\apj} {\bf
  223} (Aug., 1978) 1015--1031}.

\bibitem{1978ApJ...223.1032S}
F.~W. {Stecker}, \emph{{The cosmic $\gamma$-ray background from the
  annihilation of primordial stable neutral heavy leptons}},
  \href{http://dx.doi.org/10.1086/156336}{\emph{\apj} {\bf 223} (Aug., 1978)
  1032--1036}.

\bibitem{2018JCAP...02..005H}
M.~{H{\"u}tten}, C.~{Combet} and D.~{Maurin}, \emph{{Extragalactic diffuse
  {$\gamma$}-rays from dark matter annihilation: revised prediction and full
  modelling uncertainties}},
  \href{http://dx.doi.org/10.1088/1475-7516/2018/02/005}{\emph{\jcap} {\bf 2}
  (Feb., 2018) 005}, [\href{http://arxiv.org/abs/1711.08323}{{\tt
  1711.08323}}].

\bibitem{2015ApJ...805...33K}
V.~{Khaire} and R.~{Srianand}, \emph{{Star Formation History, Dust Attenuation,
  and Extragalactic Background Light}},
  \href{http://dx.doi.org/10.1088/0004-637X/805/1/33}{\emph{\apj} {\bf 805}
  (May, 2015) 33}, [\href{http://arxiv.org/abs/1405.7038}{{\tt 1405.7038}}].

\bibitem{2008A&A...487..837F}
A.~{Franceschini}, G.~{Rodighiero} and M.~{Vaccari}, \emph{{Extragalactic
  optical-infrared background radiation, its time evolution and the cosmic
  photon-photon opacity}},
  \href{http://dx.doi.org/10.1051/0004-6361:200809691}{\emph{\aap} {\bf 487}
  (Sept., 2008) 837--852}, [\href{http://arxiv.org/abs/0805.1841}{{\tt
  0805.1841}}].

\bibitem{2016ApJS..222....5A}
M.~{Ackermann}, M.~{Ajello}, W.~B. {Atwood}, L.~{Baldini}, J.~{Ballet},
  G.~{Barbiellini} et~al., \emph{{2FHL: The Second Catalog of Hard Fermi-LAT
  Sources}}, \href{http://dx.doi.org/10.3847/0067-0049/222/1/5}{\emph{\apjs}
  {\bf 222} (Jan., 2016) 5}, [\href{http://arxiv.org/abs/1508.04449}{{\tt
  1508.04449}}].

\bibitem{2012PhRvD..85h3007A}
M.~{Ackermann}, M.~{Ajello}, A.~{Albert}, L.~{Baldini}, J.~{Ballet},
  G.~{Barbiellini} et~al., \emph{{Anisotropies in the diffuse gamma-ray
  background measured by the Fermi LAT}},
  \href{http://dx.doi.org/10.1103/PhysRevD.85.083007}{\emph{\prd} {\bf 85}
  (Apr., 2012) 083007}, [\href{http://arxiv.org/abs/1202.2856}{{\tt
  1202.2856}}].

\bibitem{2016PhRvD..94l3005F}
M.~{Fornasa}, A.~{Cuoco}, J.~{Zavala}, J.~M. {Gaskins}, M.~A.
  {S{\'a}nchez-Conde}, G.~{Gomez-Vargas} et~al., \emph{{Angular power spectrum
  of the diffuse gamma-ray emission as measured by the Fermi Large Area
  Telescope and constraints on its dark matter interpretation}},
  \href{http://dx.doi.org/10.1103/PhysRevD.94.123005}{\emph{\prd} {\bf 94}
  (Dec., 2016) 123005}, [\href{http://arxiv.org/abs/1608.07289}{{\tt
  1608.07289}}].

\bibitem{2007PhRvD..75f3519A}
S.~{Ando}, E.~{Komatsu}, T.~{Narumoto} and T.~{Totani}, \emph{{Dark matter
  annihilation or unresolved astrophysical sources? Anisotropy probe of the
  origin of the cosmic gamma-ray background}},
  \href{http://dx.doi.org/10.1103/PhysRevD.75.063519}{\emph{\prd} {\bf 75}
  (Mar., 2007) 063519}, [\href{http://arxiv.org/abs/astro-ph/0612467}{{\tt
  astro-ph/0612467}}].

\bibitem{2012PhRvD..86f3004C}
A.~{Cuoco}, E.~{Komatsu} and J.~M. {Siegal-Gaskins}, \emph{{Joint anisotropy
  and source count constraints on the contribution of blazars to the diffuse
  gamma-ray background}},
  \href{http://dx.doi.org/10.1103/PhysRevD.86.063004}{\emph{\prd} {\bf 86}
  (Sept., 2012) 063004}, [\href{http://arxiv.org/abs/1202.5309}{{\tt
  1202.5309}}].

\bibitem{2014JCAP...11..021D}
M.~{Di Mauro}, A.~{Cuoco}, F.~{Donato} and J.~M. {Siegal-Gaskins},
  \emph{{Fermi-LAT {$\gamma$}-ray anisotropy and intensity explained by
  unresolved radio-loud active galactic nuclei}},
  \href{http://dx.doi.org/10.1088/1475-7516/2014/11/021}{\emph{\jcap} {\bf 11}
  (Nov., 2014) 021}, [\href{http://arxiv.org/abs/1407.3275}{{\tt 1407.3275}}].

\bibitem{2006PhRvD..73b3521A}
S.~{Ando} and E.~{Komatsu}, \emph{{Anisotropy of the cosmic gamma-ray
  background from dark matter annihilation}},
  \href{http://dx.doi.org/10.1103/PhysRevD.73.023521}{\emph{\prd} {\bf 73}
  (Jan., 2006) 023521}, [\href{http://arxiv.org/abs/astro-ph/0512217}{{\tt
  astro-ph/0512217}}].

\bibitem{2008JCAP...10..040S}
J.~M. {Siegal-Gaskins}, \emph{{Revealing dark matter substructure with
  anisotropies in the diffuse gamma-ray background}},
  \href{http://dx.doi.org/10.1088/1475-7516/2008/10/040}{\emph{\jcap} {\bf 10}
  (Oct., 2008) 040}, [\href{http://arxiv.org/abs/0807.1328}{{\tt 0807.1328}}].

\bibitem{2009PhRvD..80b3518F}
M.~{Fornasa}, L.~{Pieri}, G.~{Bertone} and E.~{Branchini}, \emph{{Anisotropy
  probe of galactic and extra-galactic dark matter annihilations}},
  \href{http://dx.doi.org/10.1103/PhysRevD.80.023518}{\emph{\prd} {\bf 80}
  (July, 2009) 023518}, [\href{http://arxiv.org/abs/0901.2921}{{\tt
  0901.2921}}].

\bibitem{2011MNRAS.414.2040C}
A.~{Cuoco}, A.~{Sellerholm}, J.~{Conrad} and S.~{Hannestad},
  \emph{{Anisotropies in the diffuse gamma-ray background from dark matter with
  Fermi LAT: a closer look}},
  \href{http://dx.doi.org/10.1111/j.1365-2966.2011.18525.x}{\emph{\mnras} {\bf
  414} (July, 2011) 2040--2054}, [\href{http://arxiv.org/abs/1005.0843}{{\tt
  1005.0843}}].

\bibitem{2013MNRAS.429.1529F}
M.~{Fornasa}, J.~{Zavala}, M.~A. {S{\'a}nchez-Conde}, J.~M. {Siegal-Gaskins},
  T.~{Delahaye}, F.~{Prada} et~al., \emph{{Characterization of
  dark-matter-induced anisotropies in the diffuse gamma-ray background}},
  \href{http://dx.doi.org/10.1093/mnras/sts444}{\emph{\mnras} {\bf 429} (Feb.,
  2013) 1529--1553}, [\href{http://arxiv.org/abs/1207.0502}{{\tt 1207.0502}}].

\bibitem{2013PhRvD..87l3539A}
S.~{Ando} and E.~{Komatsu}, \emph{{Constraints on the annihilation cross
  section of dark matter particles from anisotropies in the diffuse gamma-ray
  background measured with Fermi-LAT}},
  \href{http://dx.doi.org/10.1103/PhysRevD.87.123539}{\emph{\prd} {\bf 87}
  (June, 2013) 123539}, [\href{http://arxiv.org/abs/1301.5901}{{\tt
  1301.5901}}].

\bibitem{2014NIMPA.742..149G}
G.~A. {G{\'o}mez-Vargas}, A.~{Cuoco}, T.~{Linden}, M.~A. {S{\'a}nchez-Conde},
  J.~M. {Siegal-Gaskins}, T.~{Delahaye} et~al., \emph{{Dark matter implications
  of Fermi-LAT measurement of anisotropies in the diffuse gamma-ray
  background}},
  \href{http://dx.doi.org/10.1016/j.nima.2013.11.009}{\emph{\nima} {\bf 742}
  (Apr., 2014) 149--153}, [\href{http://arxiv.org/abs/1303.2154}{{\tt
  1303.2154}}].

\bibitem{2015MNRAS.447..939L}
J.~U. {Lange} and M.-C. {Chu}, \emph{{Can galactic dark matter substructure
  contribute to the cosmic gamma-ray anisotropy?}},
  \href{http://dx.doi.org/10.1093/mnras/stu2459}{\emph{\mnras} {\bf 447} (Feb.,
  2015) 939--947}, [\href{http://arxiv.org/abs/1412.5749}{{\tt 1412.5749}}].

\bibitem{2014JCAP...01..049R}
J.~{Ripken}, A.~{Cuoco}, H.-S. {Zechlin}, J.~{Conrad} and D.~{Horns},
  \emph{{The sensitivity of Cherenkov telescopes to dark matter and
  astrophysical anisotropies in the diffuse gamma-ray background}},
  \href{http://dx.doi.org/10.1088/1475-7516/2014/01/049}{\emph{\jcap} {\bf 1}
  (Jan., 2014) 049}, [\href{http://arxiv.org/abs/1211.6922}{{\tt 1211.6922}}].

\bibitem{2013APh....43....3A}
B.~S. {Acharya}, M.~{Actis}, T.~{Aghajani}, G.~{Agnetta}, J.~{Aguilar},
  F.~{Aharonian} et~al., \emph{{Introducing the CTA concept}},
  \href{http://dx.doi.org/10.1016/j.astropartphys.2013.01.007}{\emph{\ap} {\bf
  43} (Mar., 2013) 3--18}.

\bibitem{2017arXiv170907997C}
B.~S. {Acharya}, I.~{Agudo}, I.~A. {Samarai}, R.~{Alfaro}, J.~{Alfaro},
  C.~{Alispach} et~al., \emph{{Science with the Cherenkov Telescope Array}},
  {\emph{ArXiv e-prints} (Sept., 2017) \!\!},
  [\href{http://arxiv.org/abs/1709.07997}{{\tt 1709.07997}}].

\bibitem{1995PhRvD..52.4307K}
L.~{Knox}, \emph{{Determination of inflationary observables by cosmic microwave
  background anisotropy experiments}},
  \href{http://dx.doi.org/10.1103/PhysRevD.52.4307}{\emph{\prd} {\bf 52} (Oct.,
  1995) 4307--4318}, [\href{http://arxiv.org/abs/astro-ph/9504054}{{\tt
  astro-ph/9504054}}].

\bibitem{2001PhRvD..64h3003W}
B.~D. {Wandelt}, E.~{Hivon} and K.~M. {G{\'o}rski}, \emph{{Cosmic microwave
  background anisotropy power spectrum statistics for high precision
  cosmology}}, \href{http://dx.doi.org/10.1103/PhysRevD.64.083003}{\emph{\prd}
  {\bf 64} (Oct., 2001) 083003},
  [\href{http://arxiv.org/abs/astro-ph/0008111}{{\tt astro-ph/0008111}}].

\bibitem{2002ApJ...566...19K}
E.~{Komatsu}, B.~D. {Wandelt}, D.~N. {Spergel}, A.~J. {Banday} and K.~M.
  {G{\'o}rski}, \emph{{Measurement of the Cosmic Microwave Background
  Bispectrum on the COBE DMR Sky Maps}},
  \href{http://dx.doi.org/10.1086/337963}{\emph{\apj} {\bf 566} (Feb., 2002)
  19--29}, [\href{http://arxiv.org/abs/astro-ph/0107605}{{\tt
  astro-ph/0107605}}].

\bibitem{2004MNRAS.353...43P}
T.~{Poutanen}, D.~{Maino}, H.~{Kurki-Suonio}, E.~{Keih{\"a}nen} and E.~{Hivon},
  \emph{{Cosmic microwave background power spectrum estimation with the
  destriping technique}},
  \href{http://dx.doi.org/10.1111/j.1365-2966.2004.08043.x}{\emph{\mnras} {\bf
  353} (Sept., 2004) 43--58},
  [\href{http://arxiv.org/abs/astro-ph/0404134}{{\tt astro-ph/0404134}}].

\bibitem{2009PhRvD..80b3520A}
S.~{Ando}, \emph{{Gamma-ray background anisotropy from Galactic dark matter
  substructure}},
  \href{http://dx.doi.org/10.1103/PhysRevD.80.023520}{\emph{\prd} {\bf 80}
  (July, 2009) 023520}, [\href{http://arxiv.org/abs/0903.4685}{{\tt
  0903.4685}}].

\bibitem{2014MNRAS.442.1151C}
F.~{Calore}, V.~{De Romeri}, M.~{Di Mauro}, F.~{Donato}, J.~{Herpich}, A.~V.
  {Macci{\`o}} et~al., \emph{{{$\gamma$}-ray anisotropies from dark matter in
  the Milky Way: the role of the radial distribution}},
  \href{http://dx.doi.org/10.1093/mnras/stu912}{\emph{\mnras} {\bf 442} (Aug.,
  2014) 1151--1156}, [\href{http://arxiv.org/abs/1402.0512}{{\tt 1402.0512}}].

\bibitem{2016JCAP...09..047H}
M.~{H{\"u}tten}, C.~{Combet}, G.~{Maier} and D.~{Maurin}, \emph{{Dark matter
  substructure modelling and sensitivity of the Cherenkov Telescope Array to
  Galactic dark halos}},
  \href{http://dx.doi.org/10.1088/1475-7516/2016/09/047}{\emph{\jcap} {\bf 9}
  (Sept., 2016) 047}, [\href{http://arxiv.org/abs/1606.04898}{{\tt
  1606.04898}}].

\bibitem{2013APh....43..317D}
G.~{Dubus}, J.~L. {Contreras}, S.~{Funk}, Y.~{Gallant}, T.~{Hassan},
  J.~{Hinton} et~al., \emph{{Surveys with the Cherenkov Telescope Array}},
  \href{http://dx.doi.org/10.1016/j.astropartphys.2012.05.020}{\emph{\ap} {\bf
  43} (Mar., 2013) 317--330}, [\href{http://arxiv.org/abs/1208.5686}{{\tt
  1208.5686}}].

\bibitem{2016A&A...593A...1K}
J.~{Kn{\"o}dlseder}, M.~{Mayer}, C.~{Deil}, J.-B. {Cayrou}, E.~{Owen},
  N.~{Kelley-Hoskins} et~al., \emph{{GammaLib and ctools. A software framework
  for the analysis of astronomical gamma-ray data}},
  \href{http://dx.doi.org/10.1051/0004-6361/201628822}{\emph{\aap} {\bf 593}
  (Aug., 2016) A1}, [\href{http://arxiv.org/abs/1606.00393}{{\tt 1606.00393}}].

\bibitem{2005ApJ...622..759G}
K.~M. {G{\'o}rski}, E.~{Hivon}, A.~J. {Banday}, B.~D. {Wandelt}, F.~K.
  {Hansen}, M.~{Reinecke} et~al., \emph{{HEALPix: A Framework for
  High-Resolution Discretization and Fast Analysis of Data Distributed on the
  Sphere}}, \href{http://dx.doi.org/10.1086/427976}{\emph{\apj} {\bf 622}
  (Apr., 2005) 759--771}, [\href{http://arxiv.org/abs/astro-ph/0409513}{{\tt
  astro-ph/0409513}}].

\bibitem{2011EPJC...71.1554C}
G.~{Cowan}, K.~{Cranmer}, E.~{Gross} and O.~{Vitells}, \emph{{Asymptotic
  formulae for likelihood-based tests of new physics}},
  \href{http://dx.doi.org/10.1140/epjc/s10052-011-1554-0}{\emph{European
  Physical Journal C} {\bf 71} (Feb., 2011) 1554},
  [\href{http://arxiv.org/abs/1007.1727}{{\tt 1007.1727}}].

\bibitem{2001ApJ...548L.115S}
I.~{Szapudi}, S.~{Prunet}, D.~{Pogosyan}, A.~S. {Szalay} and J.~R. {Bond},
  \emph{{Fast Cosmic Microwave Background Analyses via Correlation Functions}},
  \href{http://dx.doi.org/10.1086/319105}{\emph{\apjl} {\bf 548} (Feb., 2001)
  L115--L118}.

\bibitem{2011ascl.soft09005C}
A.~{Challinor}, G.~{Chon}, S.~{Colombi}, E.~{Hivon}, S.~{Prunet} and
  I.~{Szapudi}, ``{PolSpice: Spatially Inhomogeneous Correlation Estimator for
  Temperature and Polarisation}.'' Astrophysics Source Code Library, Sept.,
  2011.

\bibitem{2002ApJ...567....2H}
E.~{Hivon}, K.~M. {G{\'o}rski}, C.~B. {Netterfield}, B.~P. {Crill}, S.~{Prunet}
  and F.~{Hansen}, \emph{{MASTER of the Cosmic Microwave Background Anisotropy
  Power Spectrum: A Fast Method for Statistical Analysis of Large and Complex
  Cosmic Microwave Background Data Sets}},
  \href{http://dx.doi.org/10.1086/338126}{\emph{\apj} {\bf 567} (Mar., 2002)
  2--17}, [\href{http://arxiv.org/abs/astro-ph/0105302}{{\tt
  astro-ph/0105302}}].

\bibitem{2015MNRAS.448.2854C}
S.~S. {Campbell}, \emph{{Angular power spectra with finite counts}},
  \href{http://dx.doi.org/10.1093/mnras/stv135}{\emph{\mnras} {\bf 448} (Apr.,
  2015) 2854--2878}, [\href{http://arxiv.org/abs/1411.4031}{{\tt 1411.4031}}].

\bibitem{2009MNRAS.400.2122A}
S.~{Ando} and V.~{Pavlidou}, \emph{{Imprint of galaxy clustering in the cosmic
  gamma-ray background}},
  \href{http://dx.doi.org/10.1111/j.1365-2966.2009.15605.x}{\emph{\mnras} {\bf
  400} (Dec., 2009) 2122--2127}, [\href{http://arxiv.org/abs/0908.3890}{{\tt
  0908.3890}}].

\bibitem{2005ApJ...626L..29A}
M.~{Amenomori}, S.~{Ayabe}, S.~W. {Cui}, {Danzengluobu}, L.~K. {Ding}, X.~H.
  {Ding} et~al., \emph{{Large-Scale Sidereal Anisotropy of Galactic Cosmic-Ray
  Intensity Observed by the Tibet Air Shower Array}},
  \href{http://dx.doi.org/10.1086/431582}{\emph{\apjl} {\bf 626} (June, 2005)
  L29--L32}, [\href{http://arxiv.org/abs/astro-ph/0505114}{{\tt
  astro-ph/0505114}}].

\bibitem{2009ApJ...698.2121A}
A.~A. {Abdo}, B.~T. {Allen}, T.~{Aune}, D.~{Berley}, S.~{Casanova}, C.~{Chen}
  et~al., \emph{{The Large-Scale Cosmic-Ray Anisotropy as Observed with
  Milagro}}, \href{http://dx.doi.org/10.1088/0004-637X/698/2/2121}{\emph{\apj}
  {\bf 698} (June, 2009) 2121--2130},
  [\href{http://arxiv.org/abs/0806.2293}{{\tt 0806.2293}}].

\bibitem{2015arXiv151004134T}
{The HAWC Collaboration} and {The IceCube Collaboration}, \emph{{Full-Sky
  Analysis of Cosmic-Ray Anisotropy with IceCube and HAWC}}, {\emph{ArXiv
  e-prints} (Oct., 2015) \!\!}, [\href{http://arxiv.org/abs/1510.04134}{{\tt
  1510.04134}}].

\bibitem{2017arXiv170803005T}
{The HAWC Collaboration} and {The IceCube Collaboration}, \emph{{Combined
  Analysis of Cosmic-Ray Anisotropy with IceCube and HAWC}}, {\emph{ArXiv
  e-prints} (Aug., 2017) \!\!}, [\href{http://arxiv.org/abs/1708.03005}{{\tt
  1708.03005}}].

\bibitem{2017PhRvL.118i1103A}
S.~{Abdollahi}, M.~{Ackermann}, M.~{Ajello}, A.~{Albert}, W.~B. {Atwood},
  L.~{Baldini} et~al., \emph{{Search for Cosmic-Ray Electron and Positron
  Anisotropies with Seven Years of Fermi Large Area Telescope Data}},
  \href{http://dx.doi.org/10.1103/PhysRevLett.118.091103}{\emph{\prl} {\bf 118}
  (Mar., 2017) 091103}, [\href{http://arxiv.org/abs/1703.01073}{{\tt
  1703.01073}}].

\bibitem{2015PhLB..749..267L}
X.~{Li}, Z.-Q. {Shen}, B.-Q. {Lu}, T.-K. {Dong}, Y.-Z. {Fan}, L.~{Feng} et~al.,
  \emph{{`Excess' of primary cosmic ray electrons}},
  \href{http://dx.doi.org/10.1016/j.physletb.2015.08.001}{\emph{Physics Letters
  B} {\bf 749} (Oct., 2015) 267--271},
  [\href{http://arxiv.org/abs/1412.1550}{{\tt 1412.1550}}].

\bibitem{2012APh....35..537B}
E.~{Borriello}, L.~{Maccione} and A.~{Cuoco}, \emph{{Dark matter electron
  anisotropy: A universal upper limit}},
  \href{http://dx.doi.org/10.1016/j.astropartphys.2011.12.001}{\emph{Astroparticle
  Physics} {\bf 35} (Mar., 2012) 537--546},
  [\href{http://arxiv.org/abs/1012.0041}{{\tt 1012.0041}}].

\bibitem{2015JCAP...02..043P}
S.~{Profumo}, \emph{{An observable electron-positron anisotropy cannot be
  generated by dark matter}},
  \href{http://dx.doi.org/10.1088/1475-7516/2015/02/043}{\emph{\jcap} {\bf 2}
  (Feb., 2015) 043}, [\href{http://arxiv.org/abs/1405.4884}{{\tt 1405.4884}}].

\bibitem{2017ifs..confE.158N}
M.~{Negro for the {\it Fermi}-LAT Collaboration}, \emph{{Study of the
  anisotropy of the unresolved gamma-ray background}},  in \emph{Proceedings of
  the 7th International Fermi Symposium}, p.~158, Oct., 2017.

\end{thebibliography}\endgroup
\end{document}